\documentclass[peerreview, draftclsnofoot, onecolumn, 12pt]{IEEEtran}		

\ifx\pdfoutput\undefined
\usepackage{graphicx}  
\else
\usepackage[pdftex]{graphicx}
\fi

				\usepackage{url}
				\usepackage[numbers,sort&compress]{natbib}
				\usepackage{array}
				\usepackage{hyperref} 				
				\usepackage{stfloats}
				\usepackage{amssymb} 
				\usepackage{acronym} 
				\usepackage{booktabs}
				\usepackage{multirow} 
				\usepackage{amsmath} 
				\usepackage{amsfonts}
				\usepackage{verbatim} 
				\usepackage[ansinew]{inputenc} 
				\usepackage[caption=false,font=footnotesize]{subfig}
				\usepackage{rotating} 
				\usepackage{amsthm} 
				\usepackage{microtype}
				\usepackage[table]{xcolor}
				\usepackage{sidecap} 
				\usepackage{wrapfig}
				\usepackage{bbding}	
				\usepackage{algorithmic} 
				\usepackage[ruled,linesnumbered]{algorithm2e}	

		\newcommand{\spc}[1]{\ensuremath\mathbb{#1}}
		\newcommand{\set}[1]{\ensuremath\mathcal{#1}}
		
			\newlength{\TABLEUP}

\begin{document}

		\title{Looking at Cellular Networks Through Canonical Domains and Conformal Mapping}

			\author{David~González~G~and~Jyri~Hämäläinen. 
			\IEEEcompsocitemizethanks{
			\IEEEcompsocthanksitem 
					David~González~G and Jyri Hämäläinen are with the Department 
					Communications and Networking, Aalto University, Finland. Corresponding email: david.gonzalezgonzalez@aalto.fi.
					}}
			\maketitle
			
			\vspace{-0.05cm}
			\begin{abstract} 
			In order to cope with the rapidly increasing service demand in cellular networks, more cells are needed with better resource usage efficiency. This poses challenges for the network planning since service demand in practical networks is not geographically uniform and, to cope with the non-uniform service demand, network deployments are becoming increasingly irregular. This paper introduces a new idea to deal with the non-uniform network topology. Rather than capturing the network character (e.g. load distribution) by means of stochastic methods, the proposed novel approach aims at transforming the analysis from the physical (irregular) domain to a canonical/dual (uniform) domain that simplifies the work due to its symmetry. To carry out this task, physical and canonical domains are connected using the conformal (Schwarz-Christoffel) mapping, that makes the rich and mature theory of Complex Analysis available. The main contribution of this paper is to introduce and validate the usability of conformal mapping in the load coupling analysis of cellular networks.
			\end{abstract}

\vspace{-0.05cm}

		
		\section{Introduction}\label{Sec:Intro}
			Due to evolution of mobile cellular networks different generations of multiple access technologies (from 1G to 4G) have been introduced~\cite{05:00013}. While it is not yet clear whether new air interface will be standardized for 5G~\cite{05:00250}, it is evident that densities of cellular networks are constantly growing to cope with the rapidly increasing service demand. Yet, geographical distribution of the service demand is not uniform, and accordingly, networks are becoming increasingly irregular. Thus, besides technical challenges on the air interface efficiency, network planning, modeling, and performance analysis are becoming increasingly difficult, yet important, research fields.\\
\indent From the perspective of network analysis, each cellular deployment is certainly `unique' in the sense that network layout, propagation conditions, and service demand are not identical in two different cellular deployments. Thus, reliable planning and analysis of the network usually takes a lot of time and computational effort since analysis is strongly dependent on the network scale and number of users (service demand volume) under consideration. Consequently, detailed system level simulations are not very suitable for network design/planning, but are more useful in performance evaluations and Radio Resource Management (RRM) testing. On the other hand, analytical approaches such as stochastic geometry~\cite{05:00184, 05:00185}, are much less costly to carry out and they have gained popularity as an alternative way to study the cellular networks. However, in most cases restrictive assumptions and heavy approximations are required, and hence, accuracy-tractability represents the fundamental trade-off between simulation-based and analytical methods.\\
\indent Our contribution is in the field of cellular network modeling and analysis. We present a novel solution framework that can be used for planning, analysis, and optimization purposes. Instead of using stochastic methods to capture the irregular nature of the network, we transfer the analysis to a dual domain that simplifies the analysis due to its symmetry. Thus, the idea aims at introducing a simple dual domain that is connected to original domain by means of a spatial transformation, more precisely, a conformal mapping~\cite{08:00076}. Let us consider the general problem of finding network deployment (base station locations and corresponding cell areas) when service demand distribution and total traffic volume are known in a certain spatial area, namely, physical domain. In the proposed approach we form a conformal (Schwarz-Christoffel) transformation from a given (polygonal) physical domain to a rectangular area, referred to as canonical domain. Finding the (deterministic) uniform network deployment in (periodic or non-periodic) canonical domain to cope with the given service volume is straightforward and can be carried out using methods known from literature. The deployment compatible with the service demand in the physical domain is then obtained by using the inverse of the original conformal mapping. \\
\indent While idea behind the proposed approach is simple, it introduces some heavy challenges: computation of conformal mappings and their corresponding inverse mappings fulfilling the requirements of our problem is not straightforward. Even more importantly, the selection of the mapping parameters such that, e.g., service demand distribution is preserved, is challenging and related to the concept of conformal equivalence. We emphasize that the proposed methodology, although being very promising, will require in future more in-depth treatment by means of classical complex analysis~\cite{08:00078}. Finally, we note that to the best of the knowledge of the authors, the idea is novel and provides a new perspective for cellular deployment studies, i.e., a new line of research. Therefore we have put more emphasis on explaining the ideas rather than technically presenting all the mathematical details, particularly those related to numerical methods.\\
The rest of the paper is organized as follows: the next section describes in more details the research problem and the proposed solution. Spatial transformations through conformal mapping are presented in Section~\ref{Sec:ConformalMapping} and in Section~\ref{Sec:CanonicalDomain} the canonical domain and the analysis therein is explained. Numerical examples and the corresponding explanations are provided in Section~\ref{Sec:NumericalExamples}. Finally, Section~\ref{Sec:ConclusionsFutWork} closes the paper with conclusions and discussion on the future work.
			
		\section{Problem Statement and Proposed Solution}\label{Sec:ProblemStatement}
\subsection{General description}\label{Sec:ProblemStatement_GD}
Cellular networks provide wireless access for users that are usually nonuniformly distributed within the network service area, by using a spatial partition of the area into cells. Traditionally, the planning has been done so that the operator is able to provide a certain Quality of Service (QoS) in terms of data rates and (high) coverage probability, i.e., users should be able to access their services and applications (almost) anytime and anywhere. Thus, the planning process starts from the characterization of the service demand. Given its importance, service demand patterns have been extensively studied in the context of cellular networks~\cite{04:00326, 05:00221}. \\
\indent In~\cite{04:00389}, a notion of \textit{irregularity} in terms of the compatibility between service demand and capacity provision was presented. In the ideal case, the network should absorb the service demand with maximum resource usage efficiency. If all serving base stations have the same amount of resources, then cell coverage areas should be planned such that cells are equally loaded. Since the demand is nonuniformly distributed, such ideal topology would also be irregular with cells of different sizes. Motivated by these simple but fundamental observations, the authors have addressed the following research problem: 

\vspace{0.2cm}
\begingroup
\leftskip2.25em
\rightskip1.25em
\noindent\textbf{Research problem}: Assume a statistical description of the service demand in the area~${\mathcal A}$, e.g., in terms of traffic distribution $\delta$ and volume $V$. Determine the spatial partition ${\mathcal A}={\mathcal A}_1\cup\dots\cup{\mathcal A}_L$ of the network service area into cells such that the service volume $V_l$ in each cell ${\mathcal A}_l$ is the same. Then select the base station locations accordingly. 
\par
\endgroup
\vspace{0.2cm}

While this problem can be addressed by various computational methods, we are proposing a novel approach based on the classical complex analysis. To that end, two domains are considered: 
\begin{itemize}
\item The {\it physical domain}, characterized by the area ${\mathcal A}$ with service demand distribution $\delta$ and volume $V$, and 
\item the {\it canonical domain}, characterized by the rectangular area ${\mathcal R}$ with uniform service demand distribution $\delta'$ and the same service volume $V$ as in the physical domain.
\end{itemize}

\begin{figure}[t]
						\centering
						\includegraphics [width = 0.98\textwidth]{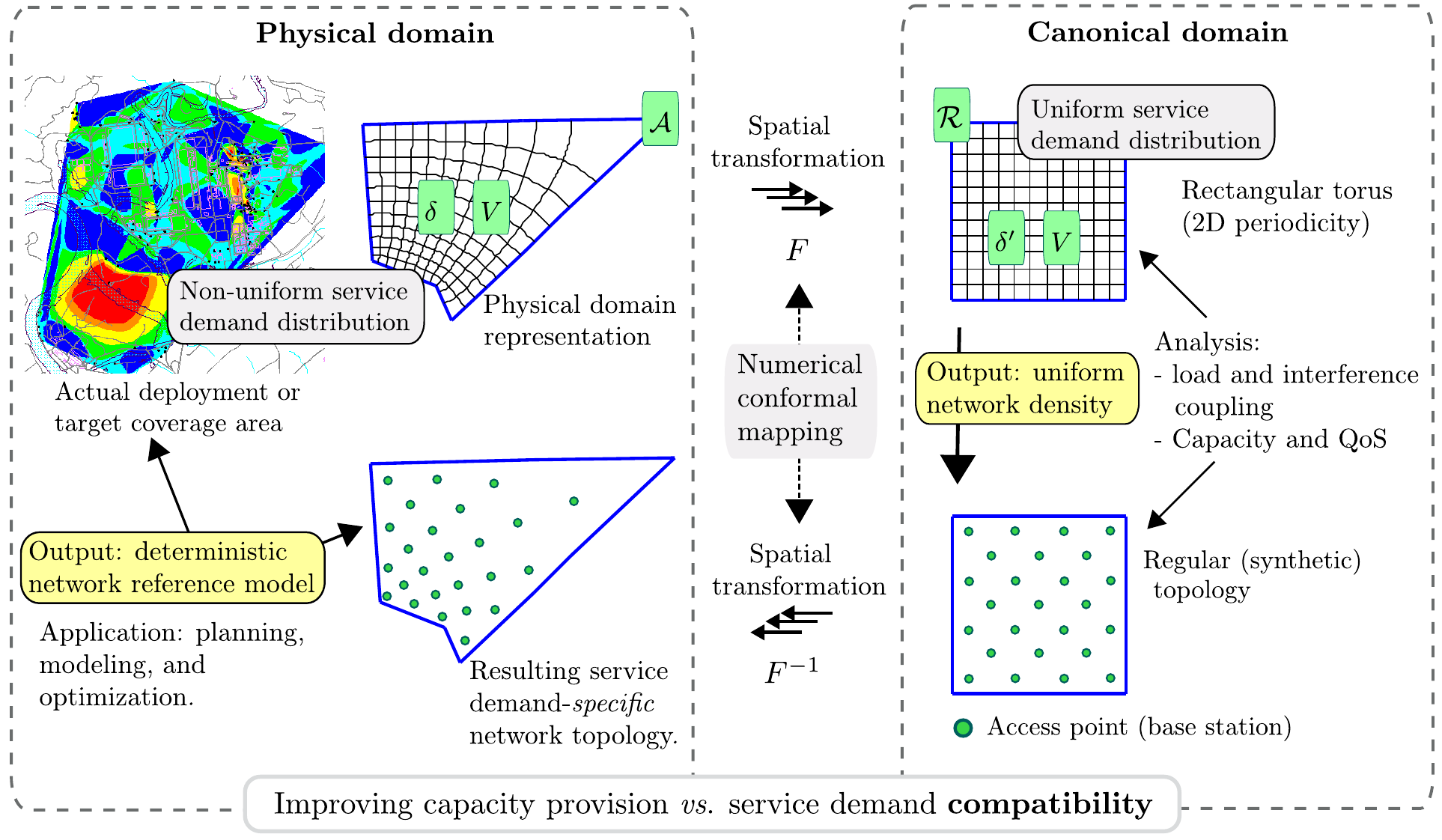}
						\vspace{-0.65cm}\caption{Research framework: Obtaining a deterministic reference model of the network that 1) is specific for the service demand spatial distribution ($\delta$) and volume ($V$), and 2) can be used for planning, load analysis, and layout/topology optimization.}
						\label{Fig:FRAMEWORK}	
					\end{figure}

In addition we need a spatial transformation $F:{\mathcal A}\rightarrow{\mathcal R}$ that preserves the load coupling analysis~\cite{05:00169} in different domains. Natural choice is to accomplish the connection between the two domains by means of \textit{conformal} mappings~\cite{08:00076}, or more precisely, by means of Schwarz-Christoffel transformations~\cite{08:00075,05:00262}. The big picture is shown in Figure~\ref{Fig:FRAMEWORK}. Details of the different phases of the solution will be provided in the next sections. 
The idea can be summarized as~follows:
\begin{enumerate}
	\item Assume a certain spatial service demand distribution $\delta$ and volume $V$. Service demand patterns are well-known, or at least good estimations are available~\cite{04:00326,05:00221}.
	\item Deduce a conformal mapping $F:{\mathcal A}\rightarrow{\mathcal R}$ such that resulting service demand distribution $\delta'$ in ${\mathcal R}$ is uniform and service demand volume is preserved.
	\item Create a network deployment in the canonical domain. That is, define the number of base stations according to the service demand volume $V$.  
	\item Deduce the inverse $F^{-1}:{\mathcal R}\rightarrow{\mathcal A}$ that is used to map the uniform base station grid from the canonical domain back onto the physical domain. 
\end{enumerate}
The phases 1) and 3) above are straightforward, however, the most challenging task in this approach is to find a suitable conformal mapping $F$ in phase 2) and its inverse $F^{-1}$ in phase 4).\\     
\indent  Finally, we note that the resulting network topology (base station locations and cell areas) provides useful information that can be used for planning, indicating the locations of the access points, or as reference in topology/energy optimization, e.g., in cases where a hyper dense network is deployed in $\mathcal{A}$. 
				
\subsection{Problem formulation and notations}\label{Sec:ProblemStatement_NotPF}
\begin{figure}[t]
						\centering
						\includegraphics [width = 0.59\textwidth]{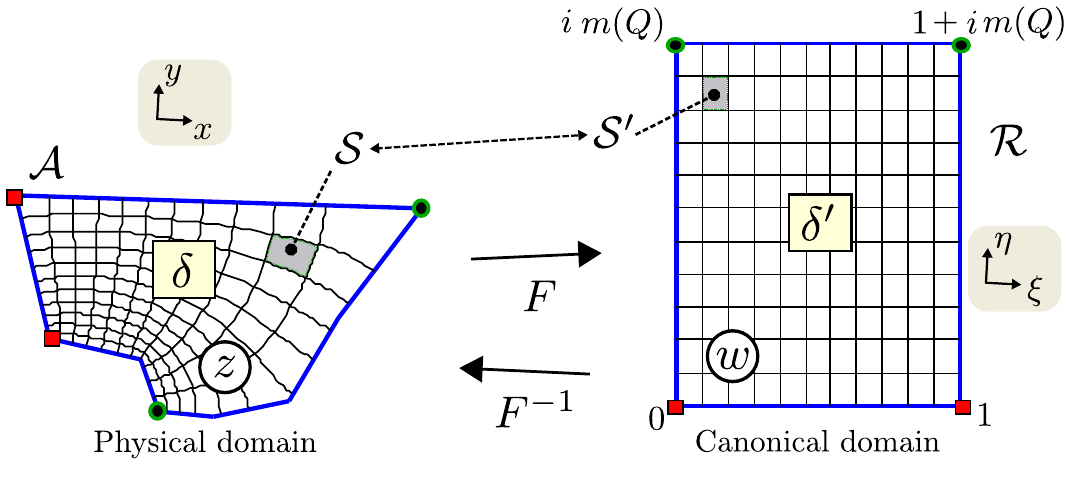}
						\vspace{-0.60cm}\caption{Representation of the required conformal mapping $F$ and its inverse $F^{-1}$. }
						\label{Fig:CONFORMAL_MAPPING_REP}	
					\end{figure}

As it was indicated, the problem at hand, that of determining the spatial mapping $F$ is difficult because several conditions need to be simultaneously fulfilled. Figure~\ref{Fig:CONFORMAL_MAPPING_REP} illustrates the required mapping indicating the complex variables $z$ and $w$ used in the physical and canonical domain, respectively. In general, the transformation $F$ and its inverse $F^{-1}$ are expressed as follows:
\begin{eqnarray}
					 &F(z)=w=u(x,y)+i\,v(x,y)=\xi+i\,\eta,\label{OP:C1}\\
						&F^{-1}(w) = z = u^{-1}(\xi,\eta)+i\,v^{-1}(\xi,\eta)=x+i\,y,\label{OP:C2}
\end{eqnarray}
where $F$ and $F^{-1}$ are conformal mappings\footnote{The notion of conformality is explained in the appendix.} if $u$, $v$,  $u^{-1}$, and  $v^{-1}$ are harmonic (analytic) functions~\cite{08:00078}.
Thus, the problem can be formulated as follows:
					\begin{subequations}
					\label{PROBLEM}
					\begin{align}					   
					\text{find}&~~F:\mathcal{A}\rightarrow\mathcal{R}~~\text{and}~~F^{-1}:\mathcal{R}\rightarrow\mathcal{A},\label{OP:MAIN}\\[-0.2cm]
					    \text{subject to:}& \nonumber\\[-0.2cm]							\textcolor[rgb]{1,1,1}{.}&\delta:\mathcal{A}\rightarrow\mathbb{R}_{+}~~\text{with}~~\int_{\mathcal{A}}\delta=1, \label{PROBLEM:C1}\\[-0.05cm]
					    \textcolor[rgb]{1,1,1}{.}&\delta^{\prime}(w)=1/m,~~\forall w\in\mathcal{R},   \label{PROBLEM:C2}\\[-0.05cm]
							\textcolor[rgb]{1,1,1}{.}&\int_{\mathcal{S}} \delta\, ds = \int_{\mathcal{S}^{\prime}} \delta^{\prime}\, ds^{\prime}.   \label{PROBLEM:C3}
					\end{align}
					\end{subequations}
					
													\begin{table}[t]
													\vspace{0.45cm}
		\caption{Notations.}
		\begin{center}
		\begin{tabular}{r l  r l }		
		\toprule  	
			 {\small \textbf{Symbol} }	& {\small\textbf{Description}}	& {\small \textbf{Symbol} }	& {\small\textbf{Description}} 			\\[-0.025cm]	 
		\midrule 			
						{\small $\mathcal{A}$ }	& {\small Canonical domain}	& {\small $\mathcal{R}$} 	& {\small Physical domain}			\\[-0.05cm]	
												{\small $\delta$ }	& {\small Spatial service demand distribution}	& {\small $\delta^{\prime}$} 	& {\small Uniform service demand distribution}			\\[-0.05cm]	
		{\small $V$} 	& {\small Service demand volume}			& {\small $V_l$} 	& {\small Service volume of the $l^{\text{th}}$ cell}		\\[-0.05cm]	
												{\small $\mu$ }	& {\small Session time}			& {\small $\lambda$ }	& {\small Inter-arrival time}	\\	[-0.05cm]					
		{\small $L$} 	& {\small Number of base stations}			&			{\small $\rho_{\text{min}}$} 	& {\small Minimum rate per user}	\\[-0.05cm]	
				 {\small $\mathcal{A}_l$ }	& {\small Area of the $l^{\text{th}}$ cell in $\mathcal{A}$.}			&	 {\small $\mathcal{R}_l$ }	& {\small Area of the $l^{\text{th}}$ cell in $\mathcal{R}$.}	\\[-0.05cm]	
				{\small $F$ }	& {\small Conformal mapping: $\mathcal{A}\rightarrow\mathcal{R}$}			& {\small $u$, $v$}	& {\small Harmonic components of $F$ }		\\[-0.05cm]	
				{\small $F^{-1}$ }	& {\small Conformal mapping: $\mathcal{R}\rightarrow\mathcal{A}$}			& {\small $u^{-1}$, $v^{-1}$ }	& {\small Harmonic components of $F^{-1}$ }	\\[-0.05cm]		
			{\small $Q$ }	& {\small Quadrilateral}			& {\small $m(Q)$ }	& {\small Conformal module of $Q$}	\\	[-0.05cm]	
						{\small $g$ }	& {\small Mapping from rectangle onto strip: (\ref{Eq:g})}			& {\small $f$ }	& {\small Mapping from rectangle onto polygon: (\ref{Eq:SC_11})}	\\	[-0.05cm]	
												{\small $H$ }	& {\small Height of the canonical domain $\mathcal{R}$}			& {\small $W$ }	& {\small Widht of the canonical domain $\mathcal{R}$}	\\	[-0.05cm]	
													{\small $\Delta^{\text{h}}$ }	& {\small Vertical shift in $\mathcal{R}$}			& {\small $\Delta^{\text{w}}$ }	& {\small Horizontal shift in  $\mathcal{R}$}	\\	[-0.05cm]																		
												{\small $d^{\text{p}}$ }	& {\small Periodic distance in $\mathcal{R}$}			& {\small $R$ }	& {\small Radius of hexagons in $\mathcal{R}$}	\\	[-0.05cm]	
												{\small $P_l$ }	& {\small Power received from the $l^{\text{th}}$ cell}			& {\small $r$ }	& {\small A given point in $\mathcal{R}$}	\\	[-0.05cm]													
												{\small $\bar{\gamma}$ }	& {\small Average SINR}			& {\small $\bar{\eta}$ }	& {\small Average spectral efficiency}	\\	[-0.05cm]			
												{\small $r^{\text{bs}}_l$ }	& {\small Location of the $l^{\text{th}}$ base station}			& {\small $\beta$ }	& {\small Propagation exponent}	\\	[-0.05cm]																																						
				\bottomrule						 
		\end{tabular} \end{center} 
		\label{Table:Not} 
		\end{table}
We note that in (\ref{PROBLEM:C2}), $m$ refers to the aspect ratio of the canonical domain and $\mathcal{S}$, $\mathcal{S}^{\prime}$ are area elements in the physical and canonical domain, respectively, as illustrated in Figure~\ref{Fig:CONFORMAL_MAPPING_REP}. In general, it is not difficult to find numerically some mappings $F$ and $F^{-1}$ between ${\mathcal A}$ and ${\mathcal R}$~\cite{05:00264}. However, Constraints (\ref{PROBLEM:C1}),~(\ref{PROBLEM:C2}),~and~(\ref{PROBLEM:C3}) make the task hard because not only any conformal mapping $F$ needs to be found, but the mapping should make the spatial service demand distribution~$\delta$ \textit{uniform} ($\delta^{\prime}$) in the canonical domain $\mathcal{R}$. In (\ref{PROBLEM:C3}), the integrations correspond to standard area integrals of the densities $\delta$ and $\delta^{\prime}$ over $\mathcal{S}$ and $\mathcal{S}^{\prime}$, respectively. Hence, $\mathcal{S}^{\prime}$ is the image of $\mathcal{S}$ after applying $F$ and $\mathcal{S}$ is the image of $\mathcal{S}^{\prime}$ after applying $F^{-1}$, as shown in Figure~\ref{Fig:CONFORMAL_MAPPING_REP}. The rationale for formulating the mapping of $\mathcal{A}$ onto a rectangle $\mathcal{R}$ will be formally justified in Section~\ref{Sec:CanonicalDomain}. Finally, in order to facilitate the reading of the rest of the article, the summary of the notations is presented in~Table~\ref{Table:Not}.			

		\section{Conformal Mapping}\label{Sec:ConformalMapping}
 In general, given a function $f(x)$ and another function $k(x,s)$, it is always possible to define the  transformation: $F(s) = \int_{\mathcal{X}}k(x,s)f(x)dx$ as long as the product $k(x,s)f(x)$ is integrable on the set $\mathcal{X}$. In this work, the use of conformal mapping, a certain spatial transformation in the complex plane, is motivated by 
the following important features:
	{\renewcommand{\labelitemi}{$\checkmark$}	
					\begin{itemize}
	\item The existence of the transformation and its inverse is guaranteed. An important and profound result from complex analysis, the Riemann's mapping theorem (see~\cite[$\text{p}.\,221$]{08:00078}) guarantees the existence of a conformal mapping between between non-empty, open, and simply connected proper subsets of $\spc{C}$, such as $\mathcal{A}$ and $\mathcal{R}$ in Figure~\ref{Fig:CONFORMAL_MAPPING_REP}. 
	\item Conformal transformations preserve notions such as locality and proximity between domains, and preserve a certain structure that is required in our context. 
					\end{itemize}}
The theory of conformal mapping is rich and mature. Comprehensive treatments can be found in~\cite{05:00264, 08:00078, 08:00076}. In any case, a brief introduction to the notion of conformality and mapping composition is provided in the appendix. Hereafter, the focus is on the type of conformal mapping used in this work: Schwarz-Christoffel transformations~\cite{08:00075}. To keep the presentation comprehensive, we explain the basics of Schwarz-Christoffel transformations in Section~\ref{Sec:ConformalMapping_SC}. In Section~\ref{Sec:ConformalMapping_P2R} we focus on mappings from polygons onto rectangles. This discussion is essential from the practical perspective since it describes the tools used to carry out phases 2) and 4) of the research problem in Section~\ref{Sec:ProblemStatement_GD}. 					
					%
\subsection{Schwarz-Christoffel~transformations: essentials and practicalities}\label{Sec:ConformalMapping_SC}
Consider the mapping shown in Figure~\ref{Fig:SC_00}. Suppose that the vertices of the polygon $\Gamma$  are $w_1,\,w_2,\,\cdots,w_n$ with the corresponding interior angles $\alpha_1\pi,\,\alpha_2\pi,\cdots,\alpha_n\pi$ and assume  that these points map onto points $z_1,\,z_2,\,\cdots,z_n$ on the real axis of the $z$-plane. Then a transformation $F$ that maps the interior of~$\Gamma$ onto the halp-plane $\mathcal{H}^{+}$ is obtained by using the Schwarz-Christoffel formula~\cite{08:00075}:
			\begin{figure}[t]
						\centering
						\includegraphics [width = 0.7\textwidth]{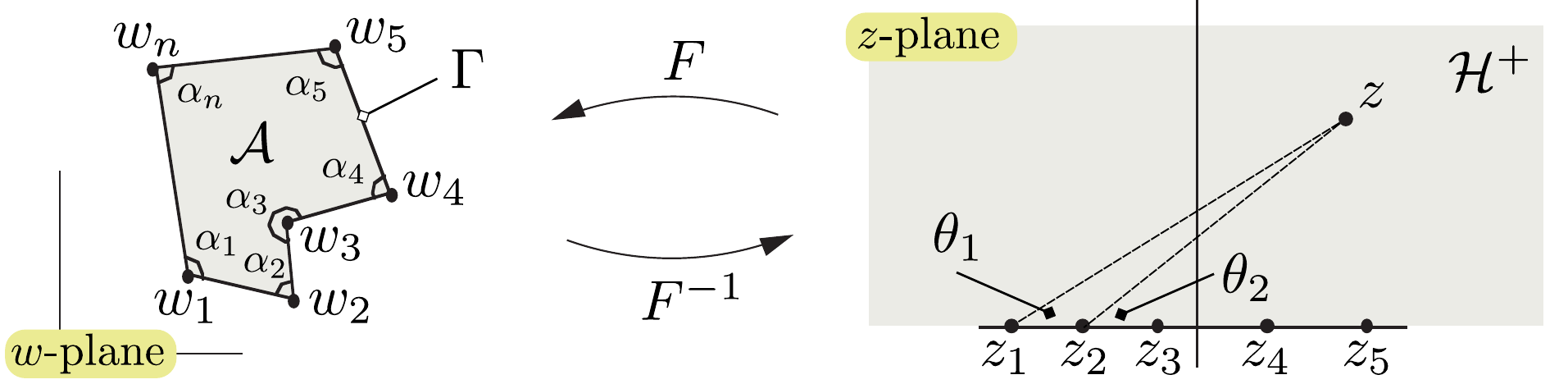}
						\vspace{-0.50cm}\caption{Schwarz-Christoffel mapping from a polygon to the half-plane ($\mathcal{H}^{+}$).}
						\label{Fig:SC_00}	
			\end{figure}
	\begin{equation}
		F(z) = A + C\,\int^z \prod_{k=1}^{n-1}\left(z - z_k\right)^{\left(\alpha_k-1\right)} dz,\label{Eq:SC_01}
	\end{equation}	
	where $A$ and $C$ are constants that adjust the size and position of the polygon ($\Gamma$) generated by the mapping and the notation $\int^z$ indicates \textsl{complex} integration~\cite{08:00078}, i.e., integration over  the complex variable $z$. The basic idea behind the Schwarz-Christoffel transformation (and its variations) is that the conformal transformation $F$ may have a derivative that can be expressed as $\frac{\partial}{\partial z}F(z) = \prod F_k$, for certain functions $F_k$~\cite{08:00075, 08:00076}. A huge variety of conformal maps admit this model, and indeed, almost all conformal transformations with known closed-form expressions belong to this family.	
	The use of Schwarz-Christoffel transformations involves three basic steps:
	\begin{enumerate}
		\item The problem of determining \textsl{the pre-vertices}, $z_1,\,z_2,\,\cdots,z_n$, which are not known \textit{a-priori}. This problem is also referred to as the Schwarz-Christoffel `parameter problem' and it is a very challenging numerical task of any Schwarz-Christoffel mapping procedure. The pre-vertices influence on the lengths of the sides in $\Gamma$. Very often, the constants $A$ and $C$ are also included in the parameter problem. 
		\item The second step involves the evaluation of the integrals in the parameter problem. Given that there are only few examples of Schwarz-Christoffel mapping that can be studied by means of closed-form expressions (typically $n\leq3$), these analytical examples are best serving as a teaching tool. Fortunately, there are well-developed numerical methods~\cite{08:00076}. We note that practical problems usually require the composition of several numerical Schwarz-Christoffel mappings, as in the case of interest herein.
		\item Finally, computational effort is needed for inverting the mapping, i.e., finding the inverse transformation. Indeed, finding the inverse mapping ($F^{-1}$) is more challenging since no formula exists in general. The vast majority of the techniques proposed so far, are based on two main approaches: Newton iteration and numerical solution of the initial value problem. A complete description of the appraches can be found in~\cite{08:00075} and references therein\footnote{This explains why the potential of Schwarz-Christoffel mapping was fully exploited much after its discovery, when computers became available. As a result, today there are several efficient numerical methods, and also a number of public domain software packages, such as~\cite{05:00265}, to assist researchers with the numerical computation burden that is required.}.
	\end{enumerate}				
\subsection{Conformal equivalence among quadrilaterals and the polygon-to-rectangle mapping}\label{Sec:ConformalMapping_P2R}
\begin{figure*}[t]
	    		\centering	  					
						\subfloat[Quadrilaterals and equivalence]
	    		{\label{Fig:FIG_CE_1}
	    		\includegraphics[width = 0.33\textwidth]{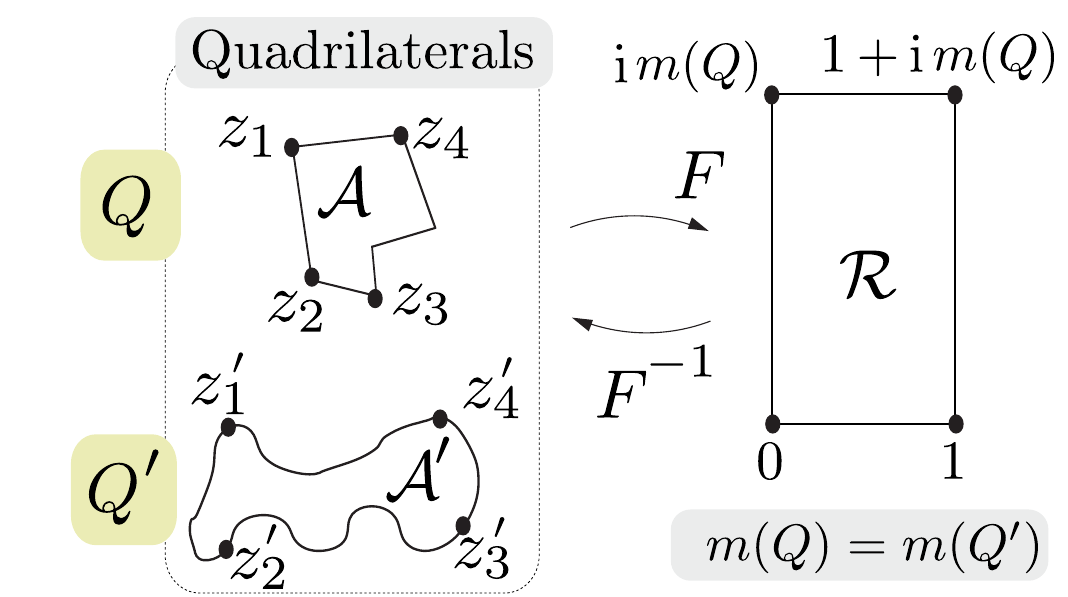}}	\hspace{0.0cm}		
															\subfloat[Transformation composition: rectangle $\leftrightarrow$ polygon]
	    		{\label{Fig:FIG_MAP_Comp}
	    		\includegraphics[width = 0.64\textwidth]{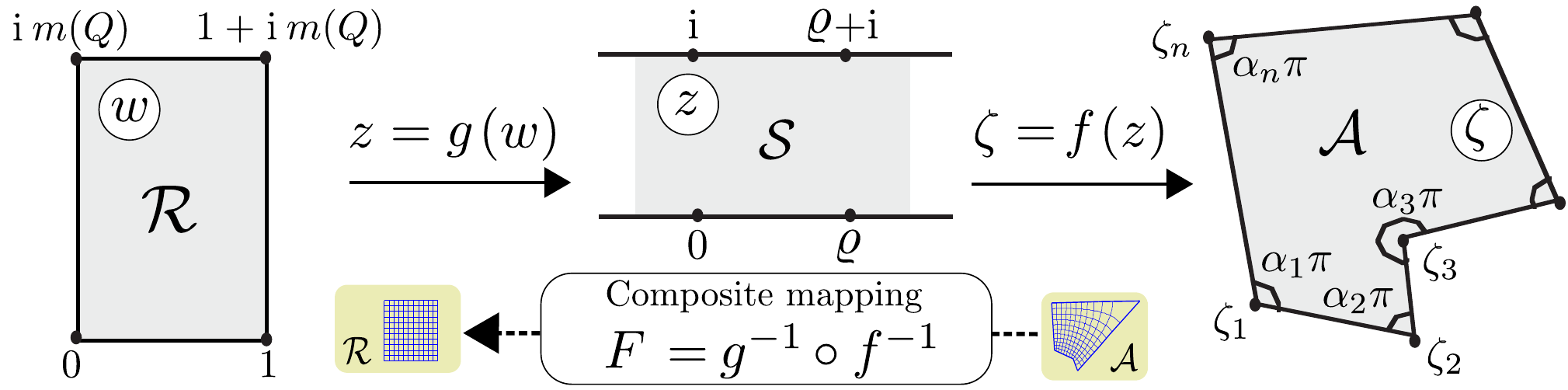}}	
	    		\vspace{-0.2cm}\caption{Conformal equivalence among quadrilaterals is a fundamental requirement for polygon-to-rectangle mapping.}
	    		\label{Fig:SC_Mod}    		
			\end{figure*}

Consider the simply-connected domains $\mathcal{A}$ and $\mathcal{A}^{\prime}$ shown in Figure~\ref{Fig:FIG_CE_1}. We note that $\mathcal{A}$ is simply-connected if a closed curve in $\mathcal{A}$ is the boundary of some area contained in $\mathcal{A}$. A \textsl{closed} curve is a curve where the beginning and end points are the same. A closed curve is called \textsl{linear} if it is a concatenation of a finite number of segments, i.e., straight lines. A curve is said to be \textsl{simple} if it does not intersect itself. A simple closed linear curve is called \textsl{polygon}. Therefore, the domain $\mathcal{A}$ is bounded by a polygon while the domain $\mathcal{A}^{\prime}$ is not bounded by a polygon, but a closed curve. A quadrilateral~\cite{08:00076} is defined as a system composed by a domain (such as $\mathcal{A}$ or $\mathcal{A}^{\prime}$) and four different points in its bounding curve. Thus, Figure~\ref{Fig:FIG_CE_1} shows two different quadrilaterals, $Q = \{\,\mathcal{A};\,z_1,\,z_2,\,z_3,\,z_4\,  \}$ and \mbox{$Q^{\prime} = \{\,\mathcal{A}^{\prime};\,z_1^{\prime},\,z_2^{\prime},\,z_3^{\prime},\,z_4^{\prime}\,  \}$}.  The `\textsl{conformal module}'~\cite{08:00076} $m(Q)$ of a quadrilateral $Q$ is defined as the unique value $m$ (see Figure~\ref{Fig:FIG_CE_1}) for which a conformal mapping between $\mathcal{A}$ and $\mathcal{R}$ exists. The conformal module $m(Q)$ is itself another unknown of the problem of determining $F:\mathcal{A}\rightarrow\mathcal{R}$. We note that the aspect ratio of $\mathcal{R}$ is completely determined by $Q$, and hence, the conformal mapping is only possible onto a rectangle $\mathcal{R}$ with a certain unique aspect ratio $m(Q)$. Two quadrilaterals $Q$ and $Q^{\prime}$ are \textsl{conformally} equivalent~\cite{08:00076}, if $m(Q)=m(Q^{\prime})$~\cite{08:00077}. Later on, the important role of $m(Q)$ will be shown when comparing the correspondence between the load coupling analysis in the canonical and physical domains.\\
\indent  Figure~\ref{Fig:FIG_MAP_Comp} illustrates the method proposed in~\cite{05:00266}, based on Schwarz-Christoffel transformations. It can be used for computing approximations to $m(Q)$ and the mapping $F^{-1}:\mathcal{R}\rightarrow \mathcal{A}$, in cases where the domain $\mathcal{A}$ is (bounded by) a polygon. The method is designed to avoid the \textit{crowding}~phenomenon~\cite[\hspace{-0.085cm}\S$\,$2.6]{08:00075}, associated with the use of previous conventional methods\footnote{The required numerical computation of the conformal map is difficult, or impossible, when the unit disc or the half plane are used as \textit{intermediate} domains~\cite{05:00266}.}, by applying an \textsl{infinite strip} (rather than the unit disc or the upper-half plane) as the intermediate domain. Thus, given a polygonal domain $\mathcal{A}$, with vertices in counterclockwise order at the points $\zeta_1,\,\zeta_2,\,\cdots,\zeta_n$, and interior angles $\alpha_1\pi,\,\alpha_2\pi,\,\cdots,\alpha_n\pi$, the method proposed in~\cite{05:00266} is based on the expression 
	 		\begin{equation}
	F^{-1} = f \circ g:\mathcal{R}\rightarrow \mathcal{A}, \label{Eq:Comp}
	\end{equation}	
	where $g:\mathcal{R}\rightarrow\mathcal{S}$ conformally maps $\mathcal{R}$ onto the infinite strip $\mathcal{S} \triangleq \{  z~|~\text{Im}(z)\in(0,1) \}$, and $f:\mathcal{S}\rightarrow\mathcal{A}$ is a modified Schwarz-Christoffel transformation. The function $g$ maps a rectangle onto a strip of width $1$, sending the corners $0$, $1$, $1+\text{i}\,m(Q)$, and $\text{i}\,m(Q)$ of the rectangle to the points $\text{i}$, $0$, $R$, and $R+\text{i}$, respectively. Then, according to~\cite{05:00266},  
		 		\begin{equation}
z=g(w)=\frac{1}{\pi}\text{log}\left[\,\text{sn}(z|m)\,\right].\label{Eq:g}
	\end{equation}	
		In~(\ref{Eq:g}), $\text{sn}(z|m)$ corresponds to the Jacobi's elliptic function~\cite{05:00264}, which is defined in terms of the incomplete elliptic integral of the first kind as follows:
			 		\begin{eqnarray}
\text{sn}(\kappa|m)&=&\text{sin}(\varphi),~~~~\text{with}\label{Eq:Jac01}\\
\kappa&=&\int_0^\varphi\frac{dt}{\sqrt{1-m\,\text{sin}^2(t)}}.\label{Eq:Jac02}
	\end{eqnarray}			
	Clearly, $\varrho$ in Figure~\ref{Fig:FIG_MAP_Comp}, is linked algebraically to the conformal module, so algebraic spacing of the points on the rectangle corresponds to algebraic spacing of the points on the strip. In this case, the conformal module is $\varrho$ and it is an unknown in the problem. Furthermore, the images of rectangle corners are constrained to lie on the vertical lines: $\text{Re}(z)=0$ and $\text{Re}(z)=R$. At this point of the overall process, there are $n-3$ unknowns, to be determined by the $n-3$ side length conditions as in the standard half-plane or disc map~\cite{08:00075}. The solution of the parameter problem implicitly determines the correct value of $R$, and hence, the elliptic parameter $m=e^{2\pi R}$ and the conformal module $m(Q)$. An appropriate formulation for solving the parameter problem is given in~\cite{05:00266}. Accordingly, the function $f$ in (\ref{Eq:Comp}) is given by
	\begin{equation}
\zeta=f(z) = A + C\,\int^z \prod_{k=1}^{n}\left[\text{sinh}\left(\frac{\pi}{2}\left(y - z_k\right)\right)\right]^{\left(\alpha_k-1\right)} dy,\label{Eq:SC_11}
	\end{equation}	
where the $z_k$'s are the pre-images on $\text{Im}(z)=0$ and $\text{Im}(z)=1$, of the \mbox{$w_k$'s, $k=1,2\cdots,n$}. Note that (\ref{Eq:SC_11}) is very similar to (\ref{Fig:SC_00}), except that the hyperbolic sines are included.\\
 \indent Thus, the composition indicated in (\ref{Eq:Comp}) is used for finding the mapping $F$ in the second step of the procedure given in Section~\ref{Sec:ProblemStatement_GD}. This approach solves problem~(\ref{PROBLEM})-(\ref{PROBLEM:C3}) in cases where the spatial service demand distribution can be expressed or approximated in terms of  analytic functions $u^{-1}$ and $v^{-1}$ that compose the mapping $F^{-1}$, see~(\ref{OP:C2}). The mathematical relationship between $F$ and $\delta$ is given by
\begin{equation}
\delta\left(z\right) = \delta\left( F^{-1}(w) \right) =K\,\frac{1}{\nabla\cdot F^{-1}} = K\,\frac{1}{\frac{\partial}{\partial \xi}\,u^{-1}\left(\xi+i\,\eta\right)+\frac{\partial}{\partial \eta}\,v^{-1}\left(\xi+i\,\eta\right)},~~~\forall\,z\in\mathcal{A},
\label{EQ:3}
\end{equation}
where $\nabla\cdot$ is the classic divergence operator and $K$ is a constant to guarantee that (\ref{PROBLEM:C1}) is fulfilled.  This implies that $\delta$ is a real analytic function. Given that (\ref{EQ:3}) is the inverse of the divergence, it can be understood as a sort of convergence operator, which is intuitively aligned with the fact that service demand is assumed to be concentrated on the area where \textit{the grid} is tighter, see Figure~\ref{Fig:CONFORMAL_MAPPING_REP}. Finally, it should be noted that the mapping $F$ and its inverse $F^{-1}$ convey information of the spatial service demand distribution $\delta$. This guarantees that the allocation of the regular network obtained in the canonical domain (discussed in the next section) is mapped onto the physical domain (through $F^{-1}$) consistently with the service demand, and more importantly, in a deterministic manner.

		\section{Analysis in the Canonical Domain}\label{Sec:CanonicalDomain}
The analysis in the canonical domain corresponds to the third step of the procedure given in Section~\ref{Sec:ProblemStatement_GD}. In what follows, we apply a load-coupling model~\cite{05:00169} to estimate the resulting uniform load in the canonical domain. However, other models such as \textit{full load}, related to the worst-case scenario, can be used as well. As indicated before, the target here is to create a regular network with a number of base stations able to satisfy the service demand volume $V$. We note that, this part of the analysis is independent of the spatial service demand distribution $\delta$ in the physical domain.

\subsection{Requirements and description}\label{Sec:CanonicalDomain_Descrip}
In~\cite{04:00389}, the notion of \textit{irregularity} is presented as a measure for the dispersion of average loads at different cells. The rationale for this concept stems from the observation that in an \textit{ideal} scenario, where the demand is uniformly distributed, a regular network topology where each cell takes exactly the same service demand and receives the same amount of intercell interference, would result in a uniformly loaded network. Hence, minimizing the irregularity, as indicated in~\cite{04:00389}, would imply that for a nonuniform demand distribution, an irregular network topology is needed to balance the load at each cell. \\ 
\indent  In order to guarantee the same load level at each cell in $\mathcal{R}$, meeting all following conditions suffices:
\begin{enumerate}
	\item The service demand share is the same for each cell, i.e., the coverage area of each cell is equal and the  demand is uniformly distributed. Hence, if there are $L$ base stations, the coverage of each cell is equal to $|\mathcal{R}_{l}| = \frac{|\mathcal{R}|}{L}$, where $|\mathcal{R}_{l}|$ is the size of the $l^{\text{th}}$ cell in $\mathcal{R}$.
	\item The distribution $f_{r}^{\text{d}}$ of the Euclidian distances in $\mathcal{R}$ is the same for all the points $r\in\mathcal{R}$.
	\end{enumerate}
	
	\indent	 The previous conditions guarantee that each cell will receive exactly the same amount of intercell interference. Intuitively, in order to fulfill the first requirement, the radio access network needs to have a regular geometry, hence, any regular tiling, e.g., rectangles or hexagons, should be used. To meet the second requirement, spatial periodicity suffices. The \textit{flat} torus provides an example of 2D manifold\footnote{A manifold is a topological space that resembles Euclidian space near each point. 2D manifolds are also called surfaces.} that is topologically equivalent to the rectangular domain $\mathcal{R}$~\cite{04:00390}. \\	
\indent	As it was indicated, the analysis in the canonical domain ($\mathcal{R}$) aims at finding the number $L$ of cells that is required to satisfy the uniformly distributed service demand with volume $V$. The analysis presented herein focuses on hexagonal tiling. The formulation for others regular tilings, such as  rectangles, can be easily inferred from the formulation presented herein. \\
\begin{figure}[t]
						\centering
						\includegraphics [width = 0.67\textwidth]{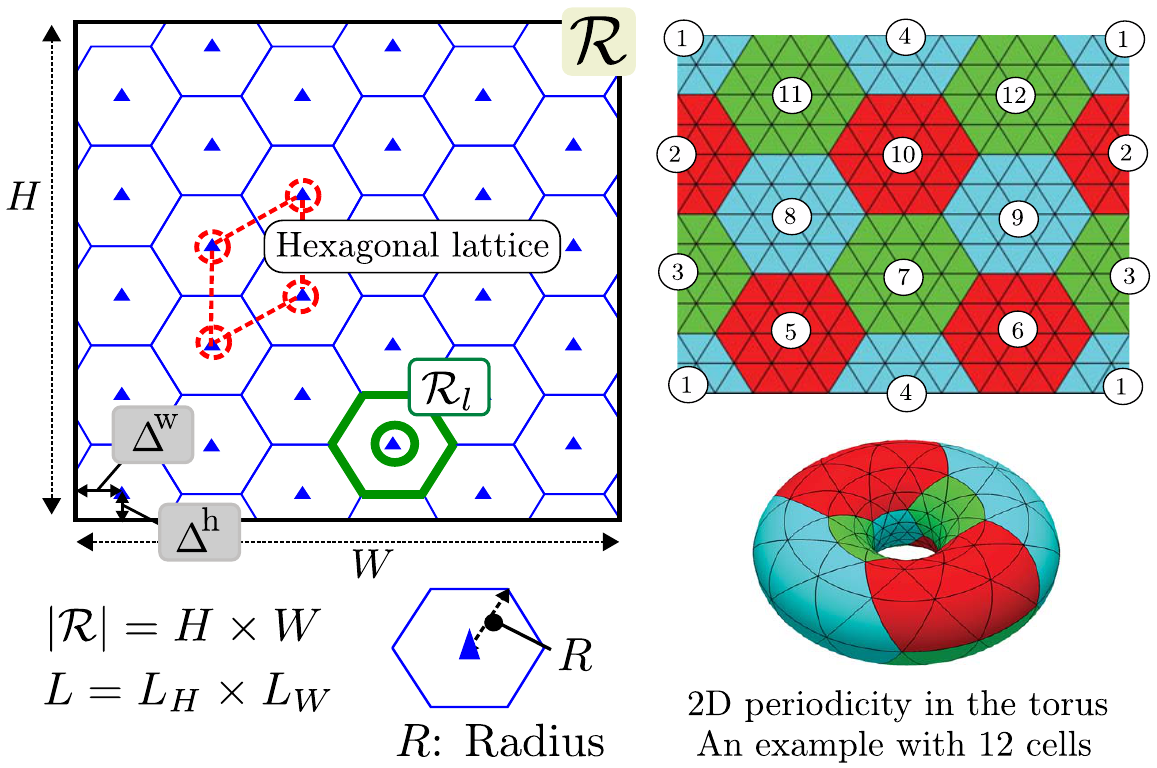}
						\vspace{-0.60cm}\caption{The canonical domain corresponds to the rectangular torus in which a regular tiling can be embedded. The torus, topologically equivalent to the rectangle, illustrates the notion of spatial  periodicity, i.e., \textit{wrap-around}.}
						\label{Fig:Canonical_Torus}	
					\end{figure}
\indent A representation of the canonical domain ($\mathcal{R}$) is shown in Figure~\ref{Fig:Canonical_Torus}. As  cells are uniformly distributed in $\mathcal{R}$, uniform density equal to $\frac{L}{H\cdot W}$ is obtained, where $W$ and $H$ are the width and height of the rectangle, respectively. The radius of each polygon is $R$, and cells  are obtained by means of the corresponding Voronoi tessellation~\cite{05:00261} that uses bases stations as generators. Thus, given $L$ access points, the $l^{\text{th}}$ cell ($\mathcal{R}_l\subset\mathcal{R}$) is defined as follows:
\begin{equation}
	\mathcal{A}_l^{\prime}  \triangleq \{\,r\in\mathcal{R} : \Vert r - r_l \Vert \leq \Vert r - r_j \Vert,~~\forall\,j\in\mathcal{L}= \{\,1,2,\cdots ,L\,\},~\,j\neq l\}.\label{Eq:Voro}
\end{equation}		
\indent In general, the allocation of an arbitrary number ($L$) of polygons in $\mathcal{R}$ is not possible. Furthermore, it is required to compute the relative angles for the selected lattice (rectangular or hexagonal), which is not a trivial task~\cite{04:00390}. In addition, as mentioned before, $\mathcal{R}$ needs to fulfill a certain aspect ratio ($\frac{H}{W}$), which cannot be controlled for some regular tilings, such as the hexagonal. Fortunately, these constraints do not represent a major issue from a practical point of view, since small deviations from the theoretical number of base stations may not imply a significant performance variation.\\
\indent Now $R =\sqrt{  \frac{2\,W\,H}{3\,\sqrt{3}\,L} }$, and obviously, $|\mathcal{R}|=W\cdot H$. Although the periodicity guarantees that the load analysis in $\mathcal{R}$ is invariant to translation, keeping the base stations as far from the boundary of $\mathcal{R}$ as possible is convenient for practical reasons. The shifts $\Delta^{\text{w}}$ and $\Delta^{\text{h}}$, see Figure~\ref{Fig:Canonical_Torus}, are given by $\frac{3\,R}{4}$ and $\frac{\sqrt{3}\,R}{4}$, respectively. Finally, after some elementary calculations, the \textit{periodic} (shortest) distance between two points~$r_1$~and~$r_2\in\mathcal{R}$ was found to be:
\begin{equation}
d^{\text{p}}(r_1,r_2)= \sqrt{\left( \frac{W}{2\,\pi}\,\text{acos}\left[ \text{cos}\left[ \left( r^{\text{w}}_{1} - r^{\text{w}}_{2}  \right)\cdot \frac{2\,\pi}{W} \right]\right]\right)^2 + \left( \frac{H}{2\,\pi}\,\text{acos}\left[ \text{cos}\left[ \left( r^{\text{h}}_{1} - r^{\text{h}}_{2}  \right)\cdot \frac{2\,\pi}{H} \right]\right]\right)^2 }.\label{Eq:DistTorus}
\end{equation}
The letters `w' and `h' refer the horizontal and vertical coordinates of $r$, $r^{\text{w}}$ and $r^{\text{h}}$, respectively.
\subsection{Load analysis}\label{Sec:CanonicalDomain_Load}
In load analysis, the target is to obtain a relationship between the network density (proportional to $L$) and the resulting uniform/average load ($\bar{\alpha}_{\text{c}}$) in the canonical domain $\mathcal{R}$. It is assumed  that there is one single type of service. Each \textit{flow} is satisfied if a minimum (average) throughput $\rho_{\text{min}}$ can be provided. In general, the demand is distributed according to some spatial distribution~$\delta:\set{X}\rightarrow\spc{R}$, where $\mathcal{X}$ is the spatial domain under consideration. Hence,
\begin{equation}
\int_{\set{X}}\delta(x)\,dx=1.\label{Eq:SpatialTrafficDist_1}
\end{equation}	
Obviously, in case of the $\mathcal{R}$, $\delta^{\prime}$ is a uniform distribution, see Figure~\ref{Fig:FRAMEWORK}. The service demand volume ($V$) is modeled in terms of two random variables: inter-arrival~($\lambda$) and session time~($\mu$). In this manner, $V$ is expressed in terms of the average number of users in the system as follows:~$V=\frac{\spc{E}\{\mu\}}{\spc{E}\{\lambda\}}$. Recall that $V$ is the same for both $\mathcal{A}$ and $\mathcal{R}$, see Figure~\ref{Fig:FRAMEWORK}.\\
\indent In general, the average Signal-to-Interference plus Noise Ratio~(SINR) at~$r\in\set{R}_{l}$ can be expressed as follows:
\begin{equation}
\bar{\gamma_{l}}(r)=  \frac{P_{l^{\star}}(r)}{\left(\displaystyle\sum\limits_{l\in\set{L}\backslash\{l^{\star}\}}\bar{\alpha}_{l}\,P_l(r) \right)+\sigma^2},  \label{Eq:SINR_Coupling_NoIntra}
\end{equation}
where $P_l(r)$ is the average received power from the $l^{\text{th}}$ base station and $\sigma^2$ is the noise power. The index~$l^{\star}$ indicates the serving base station. The instantaneous load ($\alpha_l$) at any given cell is defined as the fraction of the available bandwidth that is being used. In case of Orthogonal Frequency Division Multiple Access~(OFDMA), $\alpha_l$ is proportional to the ICI that is being generated by the cell. As it can be inferred from (\ref{Eq:SINR_Coupling_NoIntra}), the intercell interference coming from a neighbor cell is proportional to its average load ($\bar{\alpha}_l=\spc{E}\{\alpha_l\}$). In many practical cases, the SINR can be approximated by the Signal-to-Interference Ratio~(SIR), i.e., 
	\begin{equation}
\bar{\gamma_{l}}(r) \approx  \frac{P_{l^{\star}}(r)}{\displaystyle\sum\limits_{l\in\set{L}\backslash\{l^{\star}\}}\bar{\alpha}_{l}\,P_l(r) }. \label{Eq:SINR_Coupling_NoIntra2}
\end{equation}
\indent The average spectral efficiency ($\bar{\eta}$) at some point $r\in\set{R}_l$, is expressed by means of a nondecreasing function~($f_{\text{Link}}$) of the SINR, such that $\bar{\eta}({r})=f_{\text{Link}}(\,\bar{\gamma_{l}}(r)\,):\set{A}^{\prime}\rightarrow\spc{R}~\left[\frac{\text{bps}}{\text{Hz}}\right].$ Thus, the bandwidth requirement of a user located in $r\in\set{R}_l$ is given by $\bar{b_{\text{u}}}(r) = \rho_{\text{min}}/\bar{\eta}({r})~\text{[Hz]}$.\\
\indent The average load ($\bar{\alpha}_{l}$) in the $l^{\text{th}}$ cell is now given by
\begin{equation}
 \bar{\alpha}_{l}= \frac{1}{B_{\text{sys}}}\,V_{l}\,\,b_{l}, \label{Eq:NELoadStatistically_2}
\end{equation}
where,
\begin{equation}
V_{l}= \left(  \int_{\set{R}_{l}}\delta(x)\,dx \right)\frac{\spc{E}\{\mu\}}{\spc{E}\{\lambda\}}, \label{Eq:A}
\end{equation}
and
\begin{equation}
b_{l}= \int_{\set{R}_{l}}\left(\frac{\delta(x)}{\int_{\set{R}_{l}}\delta_c(w)\,dw}\right)\bar{b_{\text{u}}}(x)\,dx~~~~\text{[Hz]}. \label{Eq:B}
\end{equation}
Here $V_{l}$ and $b_{l}$ correspond to the average number of users and bandwidth consumption in the $l^{\text{th}}$ base station, respectively, and $B_{\text{sys}}$ is the system bandwidth. In order to estimate the average load vector (\mbox{$\bar{\boldsymbol{\alpha}}=[~\bar{\alpha}_1~\bar{\alpha}_2~\cdots~\bar{\alpha}_{L}~]$}), the iterative algorithm proposed in~\cite{04:00389} can be used. Yet, given the conditions of $\mathcal{R}$, it is clear that all the cells have the same load $\bar{\alpha}_{\text{c}}$. Furthermore, the terms $P_l(r)$ in~(\ref{Eq:SINR_Coupling_NoIntra})~and~(\ref{Eq:SINR_Coupling_NoIntra2}) depend on the periodic distance ($d^{\text{p}}$), in this case between the point of interest ($r$) and the location of the $l^{\text{th}}$ base station ($r^{\text{bs}}_l$), according to (\ref{Eq:DistTorus}). Thus, under the assumption that all the access points transmit the same power, (\ref{Eq:SINR_Coupling_NoIntra2}) can be written as follows:
\begin{eqnarray}
\bar{\gamma_{l}}(r) \approx  \frac{P_{l^{\star}}(r)}{\bar{\alpha}_{\text{c}}\left(\,P_{\text{T}}(r) - P_{l^{\star}}(r)\,\right)} = \frac{\left[\,d^{\text{p}}(r,r^{\text{bs}}_{l^\star})\,\right]^{-\beta}}{\bar{\alpha}_{\text{c}}\left(\,\left(\sum_{l=1}^{L}\left[\,d^{\text{p}}(r,r^{\text{bs}}_{l})\,\right]^{-\beta}\right) - \left[\,d^{\text{p}}(r,r^{\text{bs}}_{l^\star})\,\right]^{-\beta}\,\right)}, \label{Eq:SINR_CanAprox}
\end{eqnarray}
where $P_{\text{T}}(r)$ is the total received power at $r$, $P_{l^{\star}}(r)$ is the received power from the serving base station, and~$\beta\geq2$ is the propagation exponent. The location of the serving base station ($r^{\text{bs}}_{l^\star}$) depends on the selection of~$\mathcal{R}_{l}$, see Figure~\ref{Fig:Canonical_Torus}, but in any case, $d^{\text{p}}(r,r^{\text{bs}}_{l^\star})\in(0,R]$. Since $\set{R}^{\prime}$ satisfies the conditions given in Section~\ref{Sec:CanonicalDomain_Descrip}, the complete statistic of the SINR can be obtained by analyzing the points within any of the $L$ fundamental parallelograms (one of them is indicated in dashed-red in Figure~\ref{Fig:Canonical_Torus}) by means of the formulation just given. Alternatively, analyzing the coverage of one single cell also suffices. Thus, $\bar{\alpha}_{\text{c}}$ can be obtained by analyzing only the points~$r\in\mathcal{R}_{l}$. Since the sum of the distances, and hence the received power, are independent of the any translation of the grid of access points, it is possible to write
\begin{eqnarray}
P_{\text{T}}(r) = \sum_{h=1}^{L_H}\sum_{w=1}^{L_W} \left(\left( \frac{W}{2\,\pi}\,\text{acos}\left[ \text{cos}\left(  d_{\text{w}} \,\frac{2\,\pi}{W} \right)\right]\right)^2 + \left( \frac{H}{2\,\pi}\,\text{acos}\left[ \text{cos}\left( d_{\text{h}} \,  \frac{2\,\pi}{H} \right)\right]\right)^2\right)^{-\frac{\beta}{2}},\label{Eq:PTot}
\end{eqnarray}
where it is assumed that $L=L_W\times L_H$. The terms $d_{\text{w}}$ and $d_{\text{h}}$ in (\ref{Eq:PTot}) are indeed functions of the $w$, $h$, and $r$. They are defined as follows:
\begin{eqnarray}
d_{\text{w}}(w,r) &=&   r^{\text{w}}\,-\,\left((w-1)\,\frac{3}{2}\,R\,+\,\Delta^{\text{w}}\right) , \label{PTothex2}\\[-0.1cm]
d_{\text{h}}(w,h,r) &=&   r^{\text{h}}\,-\,\left((h-1)\,\sqrt{3}\,R\,+\,\text{mod}(w+1,2)\frac{\sqrt{3}}{2}\,R\,+\,\Delta^{\text{h}}\right)\label{PTothex3}.
\end{eqnarray}
Altogether, assuming $f_{\text{Link}}(x)=\text{log}_2(1+x)$, the relationship between the resulting load $\bar{\alpha}_{\text{c}}$ and the network density (or equivalently $L$) will be given by the following expression:
\begin{equation}
 L = \left\lceil\frac{r_{\text{min}}}{B_{\text{sys}}} \frac{\spc{E}\{\mu\}}{\spc{E}\{\lambda\}}\frac{1}{|\mathcal{R}|\cdot \bar{\alpha}_{\text{c}}} \int\limits_{\mathcal{R}_{l}} \frac{1}{\text{log}_2\left(1+ \frac{P_{l^{\star}}(r)}{\bar{\alpha}_{\text{c}}\cdot\left[\,P_{\text{T}}(r) - P_{l^{\star}}(r)\,\right]}\right)}\,dr\right\rceil, \label{Eq:ALPHA_L_1}
\end{equation}
where $\lceil\cdot\rceil$ refers to rounding up to the closets feasible integer.

		\section{Numerical Examples}\label{Sec:NumericalExamples}
To illustrate the proposed methodology, some numerical examples are provided. First, we introduce two example domains, related service demand distributions, and discuss the computation of the required transformation. In Section~\ref{Sec:NumericalExamples_ConDom}, we carry out the required load analysis in the canonical domain and in Section~\ref{Sec:NumericCorres} we discuss the correspondence between the canonical and physical domains in the light of examples.
\subsection{Conformal mapping}\label{Sec:NumericalExamples_ConfMapp}
\begin{figure*}[t]
	    		\centering	    		
							\subfloat[Physical domain 1: $\mathcal{A}_1$]
	    		{\label{Fig:FIG3_Phy02}
	    		\includegraphics[width = 0.28\textwidth]{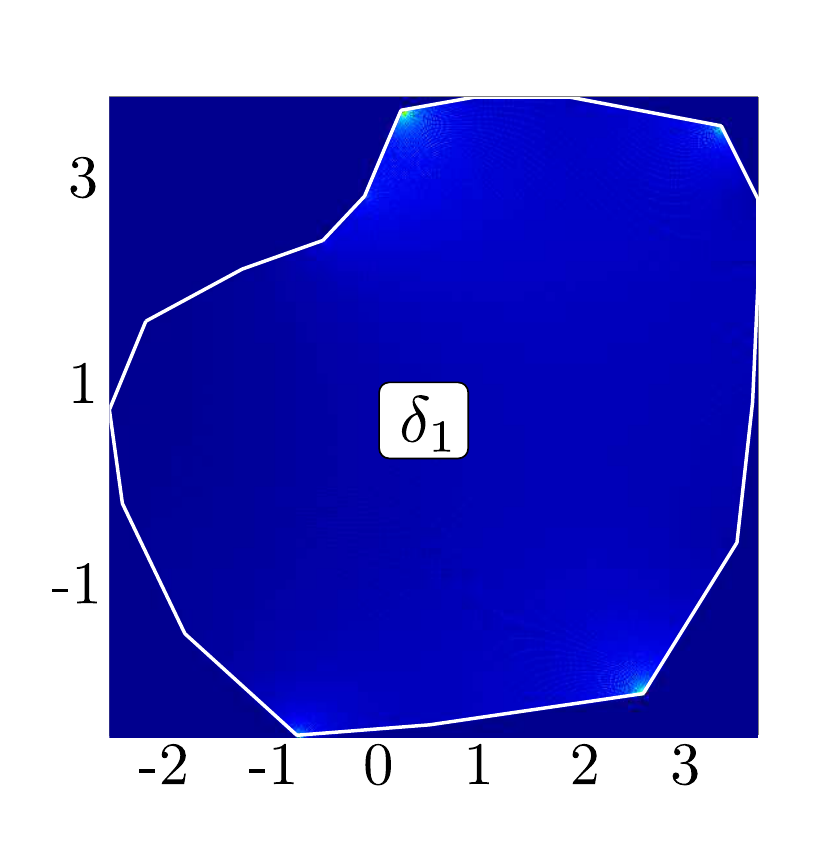}}\hspace{0.1cm}
									\subfloat[Physical domain 2: $\mathcal{A}_2$]
	    		{\label{Fig:FIG3_Phy03}
	    		\includegraphics[width = 0.28\textwidth]{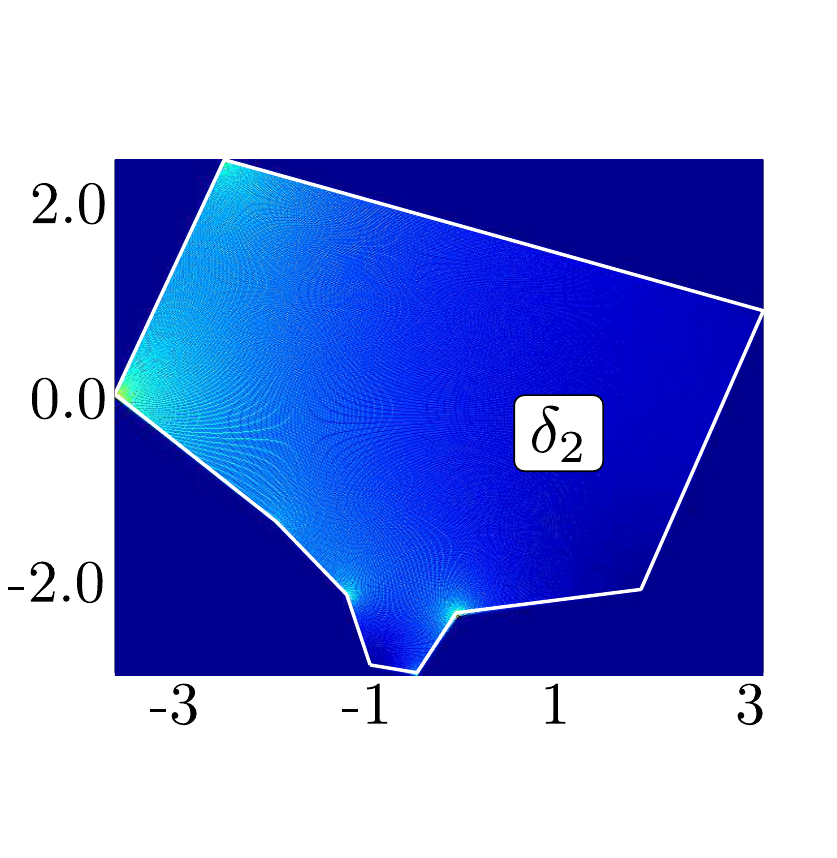}}\hspace{0.1cm}
							\subfloat[CDFs: spatial service demand]
	    		{\label{Fig:FIG3_CDFs}
	    		\includegraphics[width = 0.24\textwidth]{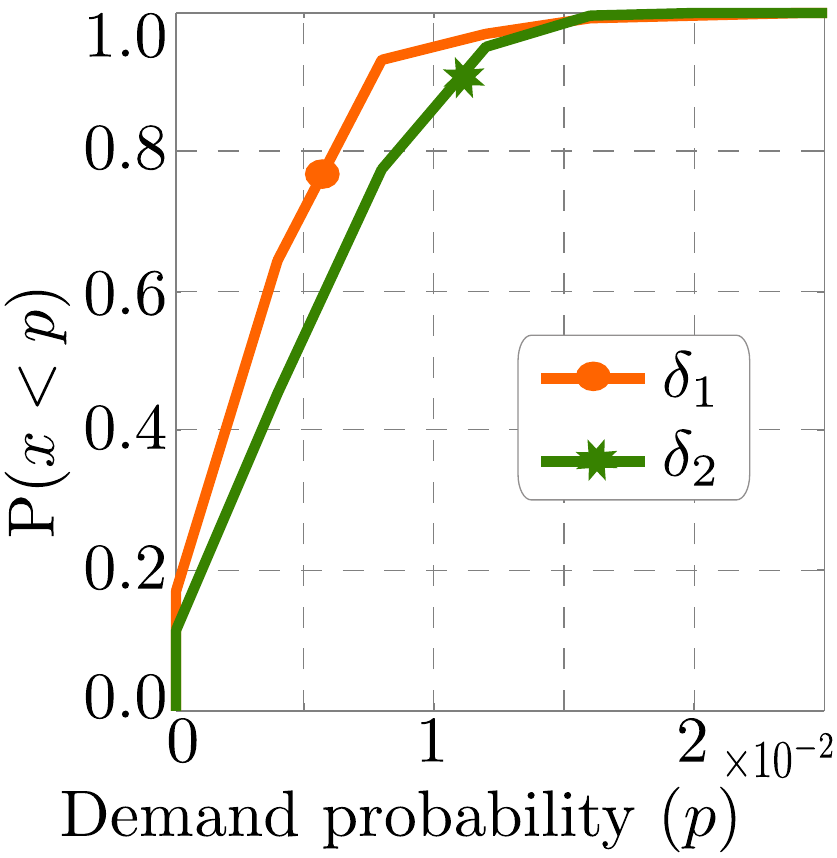}}			
	    		\vspace{-0.25cm}\caption{Physical domains considered for numerical examples.}
	    		\label{Fig:FIG_03}    		
			\end{figure*}

\begin{figure*}[t]
	    		\centering	    		
							\subfloat[Conformal mapping $\mathcal{A}_1\leftrightarrow\mathcal{R}_1$]
	    		{\label{Fig:FIG4_Phy02}
	    		\includegraphics[width = 0.35\textwidth]{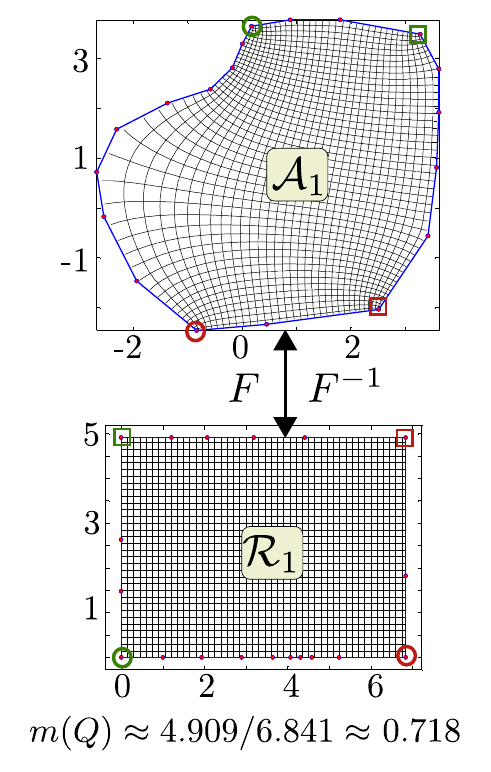}}
									\subfloat[Conformal mapping $\mathcal{A}_2\leftrightarrow\mathcal{R}_2$]
	    		{\label{Fig:FIG4_Phy03}
	    		\includegraphics[width = 0.35\textwidth]{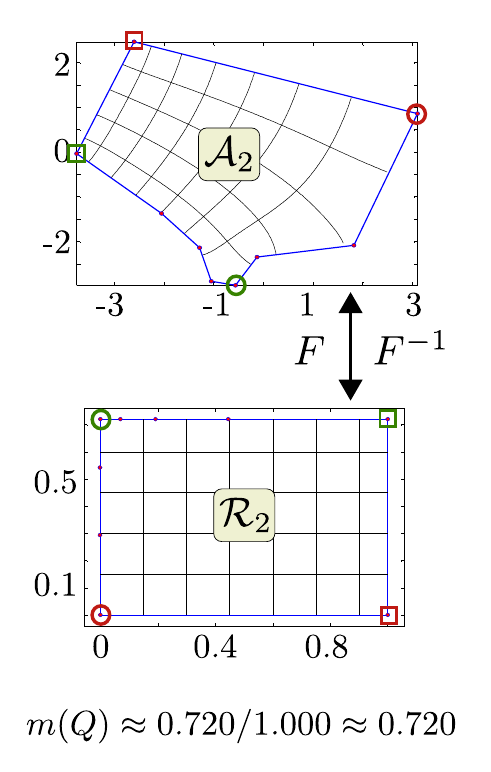}}		
	    		\vspace{-0.2cm}\caption{Representation of the Schwarz-Christoffel mappings used as examples.}
	    		\label{Fig:FIG_04}    		
			\end{figure*}

In order to illustrate aspects of interest, two different physical domains have been considered for numerical evaluations. The domains $\mathcal{A}_1$ and $\mathcal{A}_2$ are shown in Figures~\ref{Fig:FIG3_Phy02}~and~\ref{Fig:FIG3_Phy03}, respectively. As it can be seen, both domains correspond to simple linear curves, i.e., polygons, although with different number of vertices. Moreover, nonuniform spatial service demand  distributions ($\delta_1~\text{and}~\delta_2$) are defined in each case. The Cumulative Distribution Function~(CDF) of demand probability (considering $500\times500$ area elements in each case) is also shown in Figure~\ref{Fig:FIG3_CDFs}.\\ 
\indent Numerical composite mappings~\cite{08:00075}, as described in Section~\ref{Sec:ConformalMapping_P2R}, are used to redistribute these spatial service demand onto canonical domains (rectangles). Mappings are obtained by using the Schwarz-Christoffel software provided in~\cite{05:00265}. Spatial representations of applied mappings are illustrated in Figures~\ref{Fig:FIG4_Phy02}~and~\ref{Fig:FIG4_Phy03}. The vertices and prevertices of the composite Schwarz-Christoffel transformation are indicated in the boundary of each domain by small pink circles. The 4 \textit{corners} of each quadrilateral (and their images) are indicated by larger red and green circles and squares. Note that, as it was remarked earlier, the conformal module,~$m(Q)$, is an important unknown to be found. In other words, the aspect ratio of the rectangle representing the canonical domain cannot be selected arbitrarily once the quadrilateral is fixed. However, the selection of the vertices and quadrilateral's corners in the physical domain gives the degrees of freedom to find an appropriate mapping. This task needs to be carried out numerically and no exact method exists. In case of a rectangular tiling, matching the aspect ratio of $\mathcal{R}$ and $m(Q)$ can be done easily, while hexagonal lattices have their own limitations~\cite{04:00390} and ratio matching is not always possible. 
	\subsection{Load analysis in the canonical domain}\label{Sec:NumericalExamples_ConDom}	
\begin{figure}[t]
						\centering
						\includegraphics [width = 0.62\textwidth]{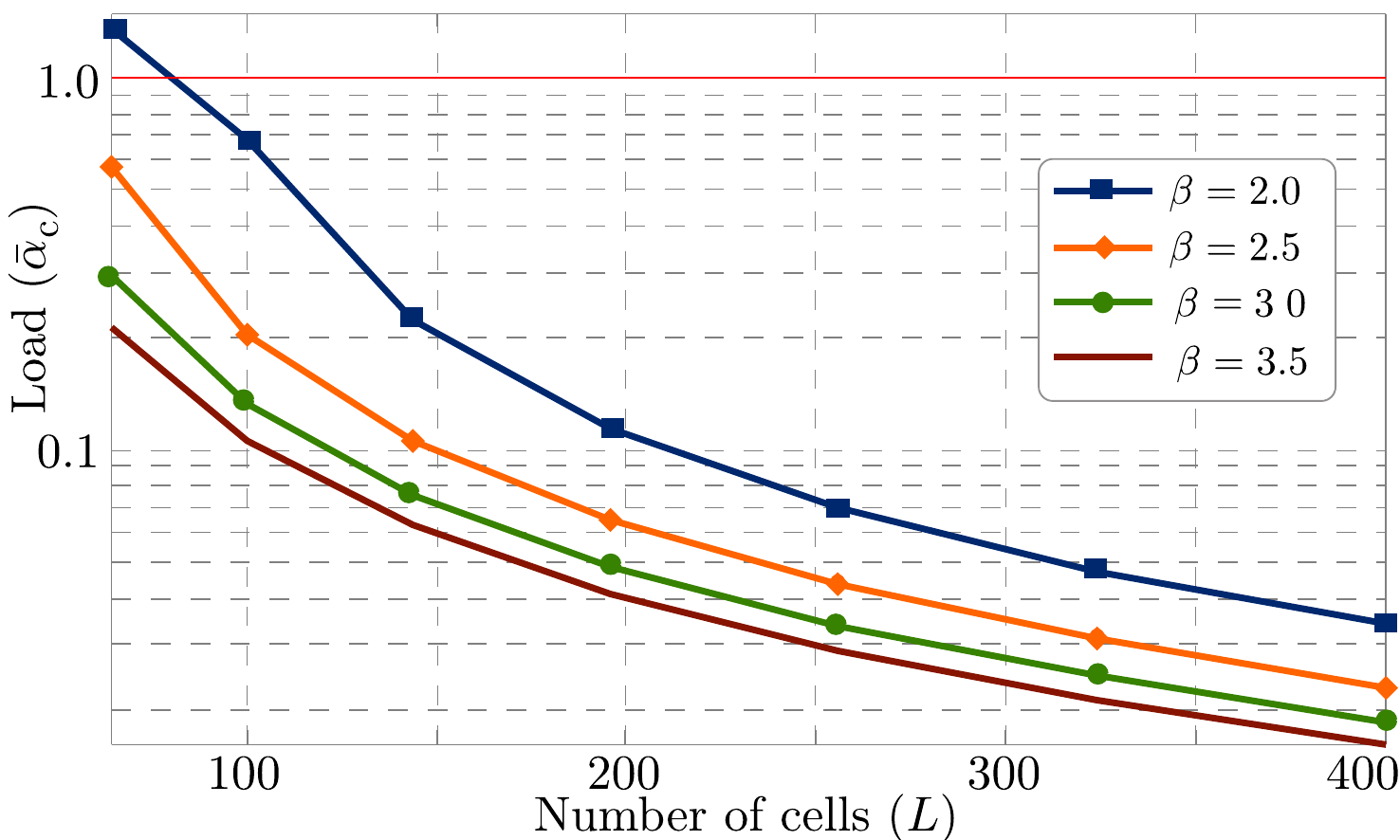}
						\vspace{-0.65cm}\caption{Load in the canonical domain (hexagonal tiling). Setting: $|\mathcal{R}|=6.84\times4.90\approx33.58$, $\spc{E}\{\mu\}=120\,\text{s}$, $\spc{E}\{\lambda\}=0.05\,\text{s}$, $L = [64,\,100,\,144,\,196,\,256,\,324,\,400]$, $B_{\text{sys}}=5.0\,\text{MHz}$, $r_{\text{min}}=100\,\text{kbps}$.}
						\label{Fig:FIG_01}	
					\end{figure}

			Figure~\ref{Fig:FIG_01} illustrates the load in the canonical domain ($\mathcal{R}$) for different  propagation exponents~($\beta$).  The vertical axis indicates the resulting uniform load $\bar{\alpha}_{\text{c}}$ as a function of number of base stations~$L$, shown in the horizontal axis.  Note that the vertical axis is given  in logarithmic scale. The figure shows that, increasing the propagation exponent results in less ICI, and hence, lower load levels, i.e., more capacity. Results of Figure~\ref{Fig:FIG_01}~are obtained by solving~(\ref{Eq:ALPHA_L_1}) for the setting indicated in the caption text. In general, the resulting load $\bar{\alpha}_{\text{c}}$ depends on the service demand volume~\cite{04:00389}, the aspect ratio of $\mathcal{R}$ (that is a constraint from the conformal mapping), and the type the of regular tiling that is used~\cite{04:00390}. However,~(\ref{Eq:ALPHA_L_1}) always allows for establishing the relation among the variables of interest ($\bar{\alpha}_{\text{c}}$, $V$, and $L$) for any rectangular domain fulfilling the conditions given in Section~\ref{Sec:CanonicalDomain_Descrip}, and hence, it is a gateway to get a relationship similar to the one shown in Figure~\ref{Fig:FIG_01} for any arbitrary physical domain through its corresponding canonical counterpart.
			\subsection{Correspondence between the canonical and physical domains}\label{Sec:NumericCorres}
\begin{figure*}[t]
	    		\centering	 
						\subfloat[Site mapping: $\mathcal{A}_1\rightarrow\mathcal{R}_1$: hexagonal tiling ($L=36$)]
	    		{\label{Fig:FIG5_Phy02hex}
	    		\includegraphics[width = 0.45\textwidth]{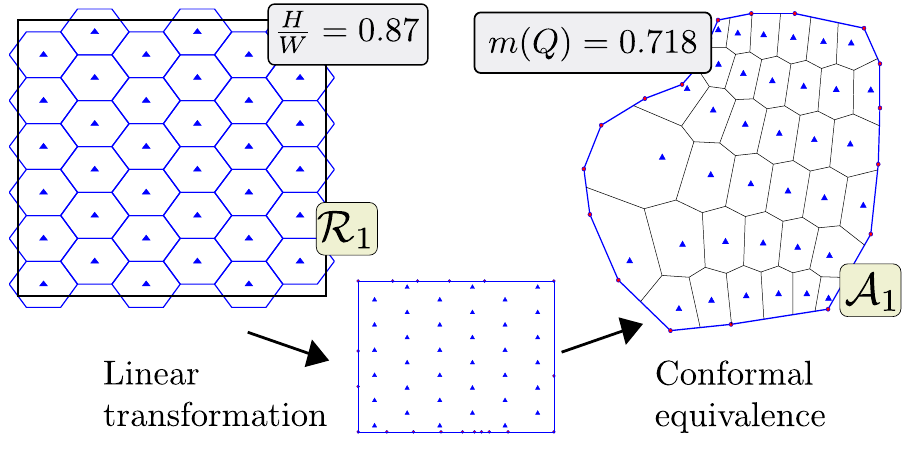}}	\hspace{0.3cm}
												\subfloat[Site mapping: $\mathcal{A}_1\rightarrow\mathcal{R}_1$: rectangular tiling ($L=36$)]
	    		{\label{Fig:FIG5_Phy02sq}
	    		\includegraphics[width = 0.45\textwidth]{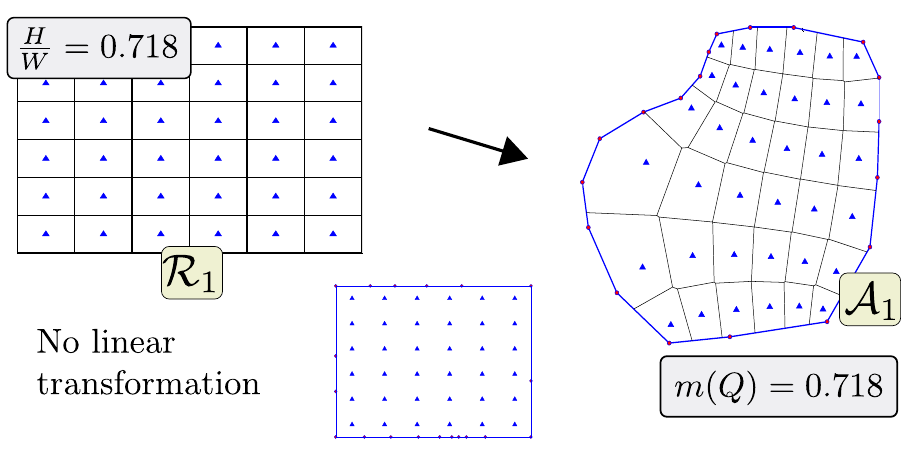}}\\						
							\subfloat[Site mapping: $\mathcal{A}_2\rightarrow\mathcal{R}_2$: hexagonal tiling ($L=180$)]
	    		{\label{Fig:FIG5_Phy03hex}
	    		\includegraphics[width = 0.45\textwidth]{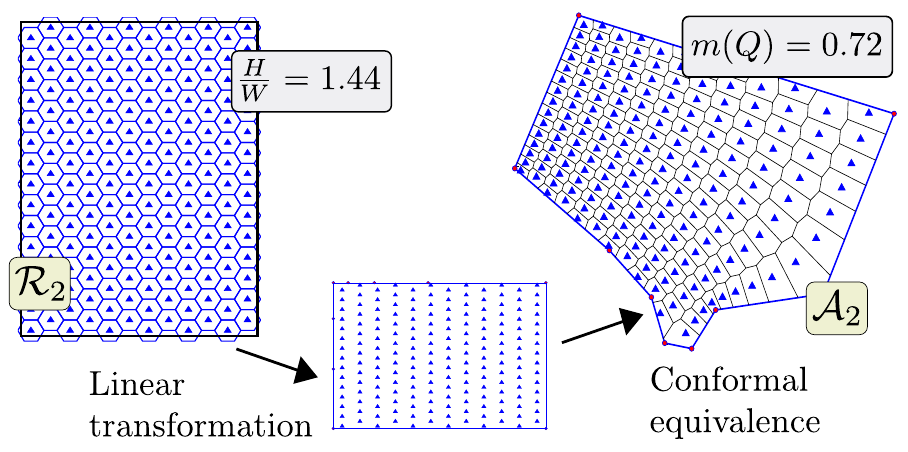}}	\hspace{0.3cm}
					\subfloat[Site mapping: $\mathcal{A}_2\rightarrow\mathcal{R}_2$: rectangular tiling ($L=180$)]
	    		{\label{Fig:FIG5_Phy03sq}
	    		\includegraphics[width = 0.45\textwidth]{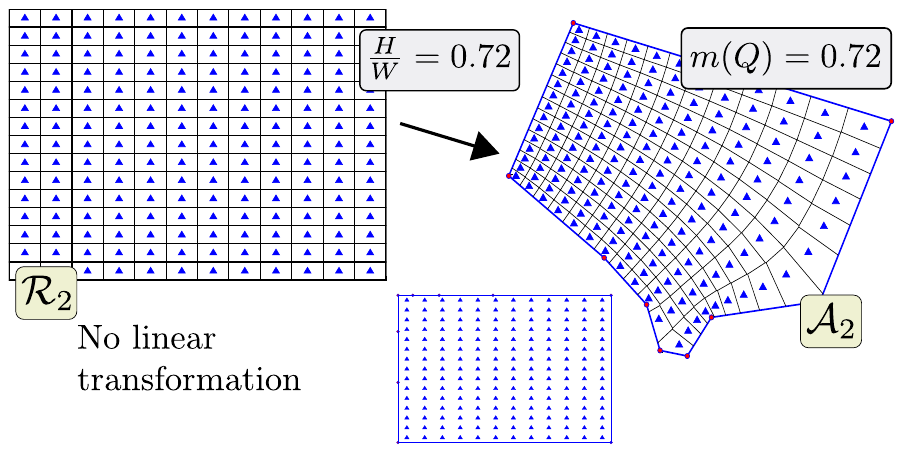}}
	    		\vspace{-0.2cm}\caption{Mapping site's locations from the canonical domain to the physical domain using  Schwarz-Christoffel transformations.}
	    		\label{Fig:FIG_SiteMapp}    		
			\end{figure*}

\begin{figure*}[t]
	    		\centering	    		
					\subfloat[$\mathcal{A}_1$-$\mathcal{R}_1$: hexagonal tiling]
	    		{\label{Fig:FIG12_1}
	    		\includegraphics[width = 0.24\textwidth]{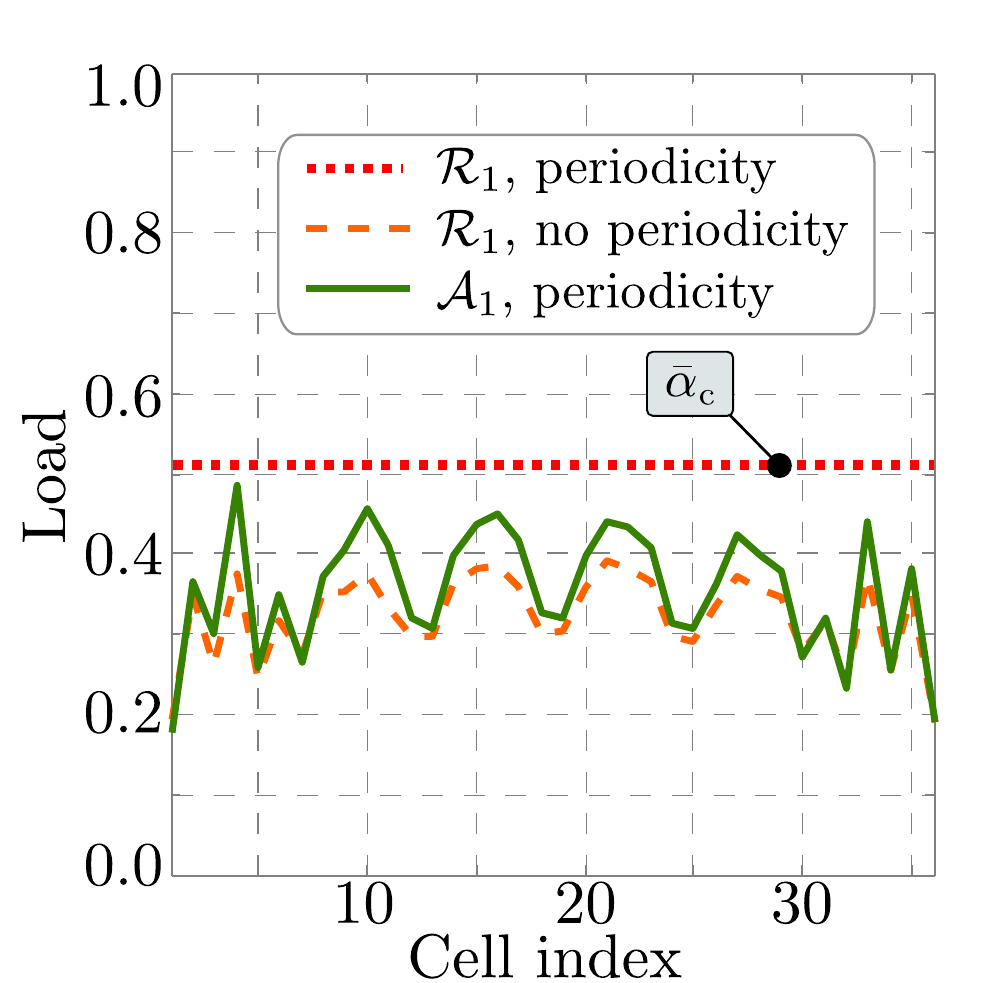}}	
					\subfloat[$\mathcal{A}_1$-$\mathcal{R}_1$: rectangular tiling]
	    		{\label{Fig:FIG12_2}
	    		\includegraphics[width = 0.24\textwidth]{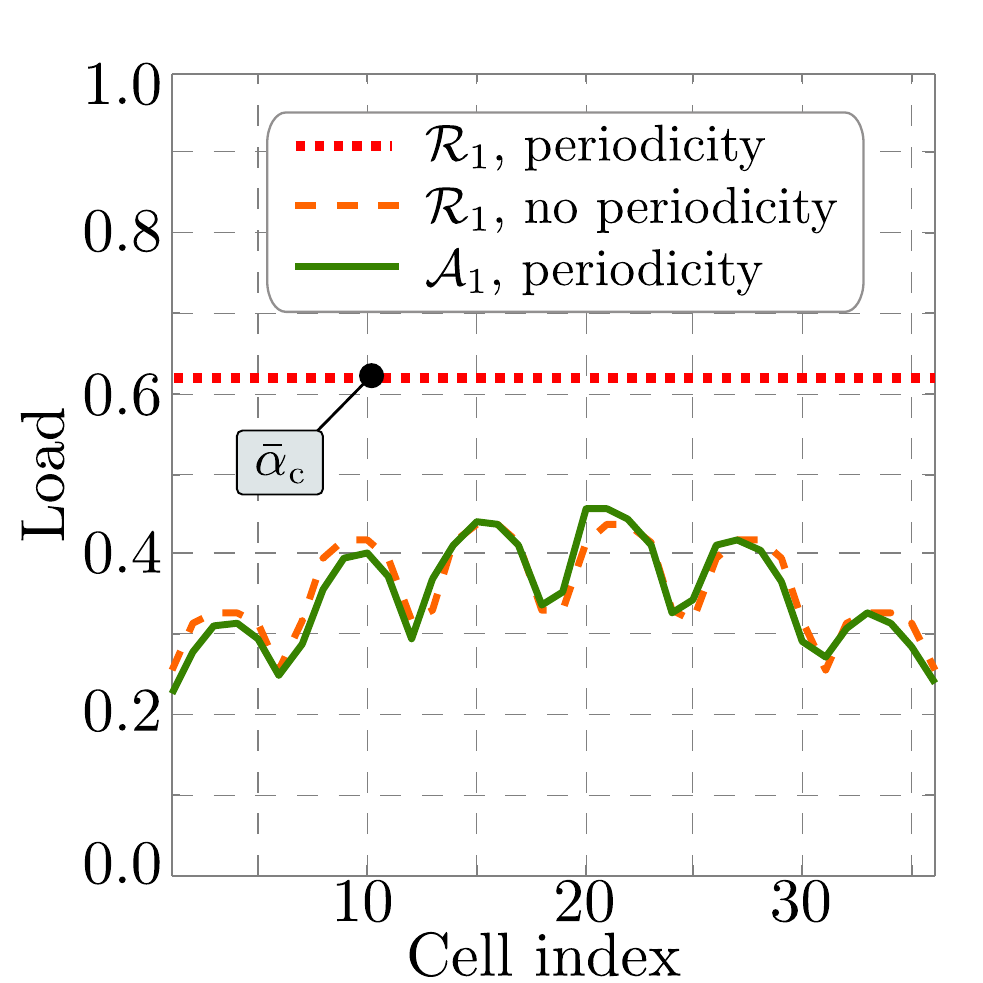}}	
					\subfloat[$\mathcal{A}_2$-$\mathcal{R}_2$: hexagonal tiling]
	    		{\label{Fig:FIG12_3}
	    		\includegraphics[width = 0.236\textwidth]{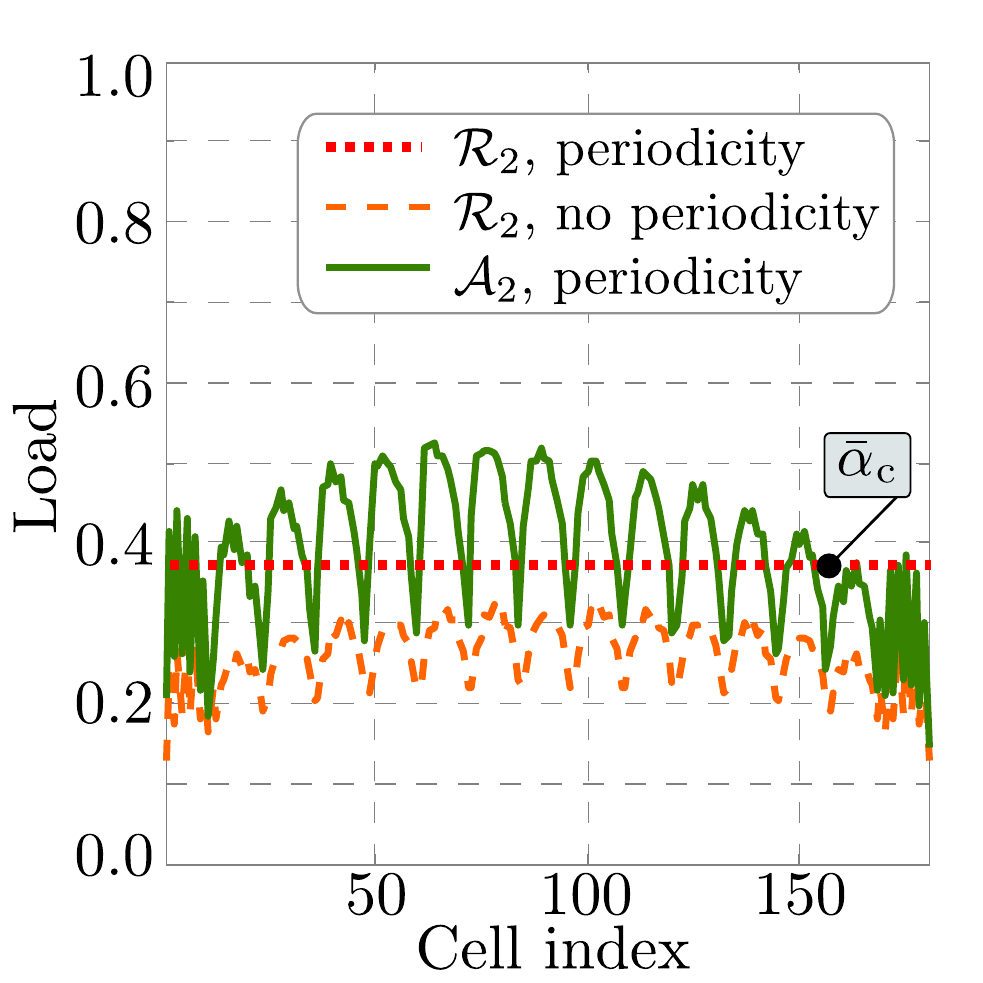}}	
					\subfloat[$\mathcal{A}_2$-$\mathcal{R}_2$: rectangular tiling]
	    		{\label{Fig:FIG12_4}
	    		\includegraphics[width = 0.236\textwidth]{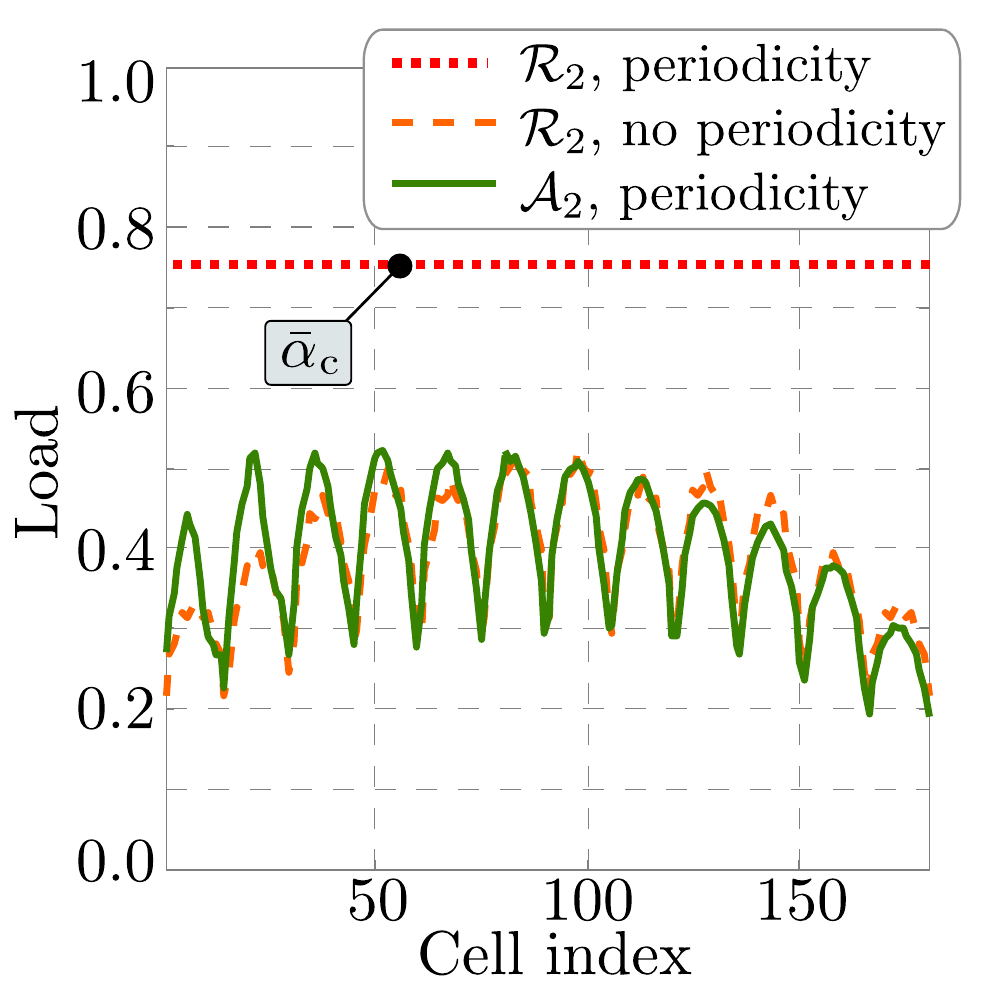}}	\\		
					\subfloat[Spatial pattern: $\mathcal{A}_1$@hex]
	    		{\label{Fig:FIG12_5}
	    		\includegraphics[width = 0.241\textwidth]{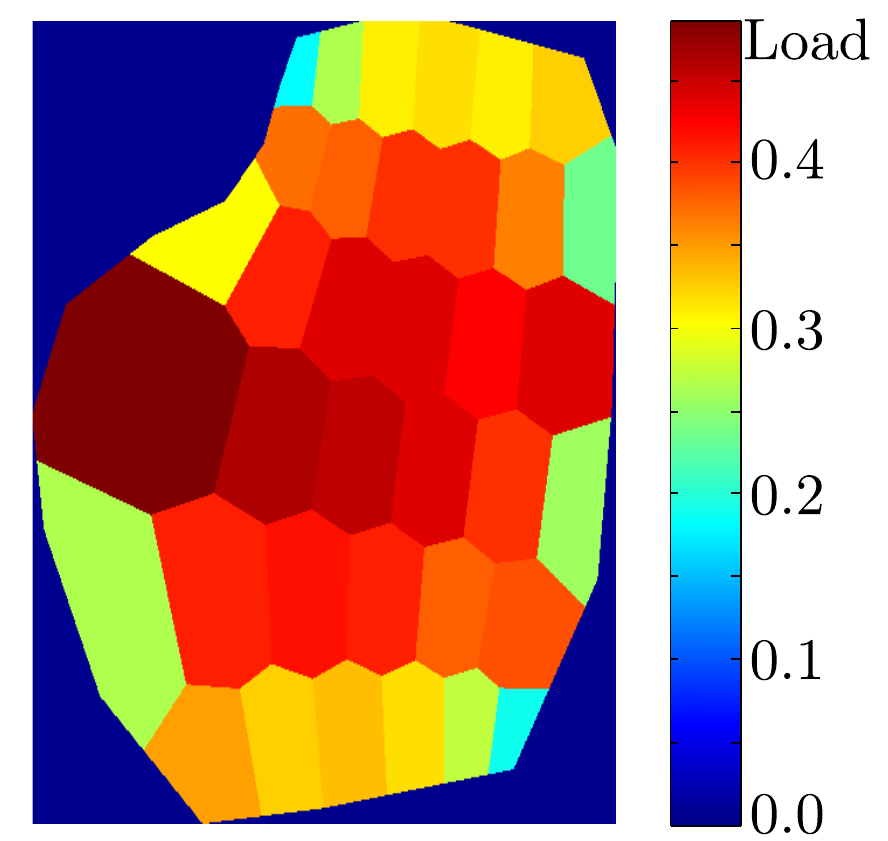}}	
					\subfloat[Spatial pattern: $\mathcal{A}_1$@rec]
	    		{\label{Fig:FIG12_6}
	    		\includegraphics[width = 0.241\textwidth]{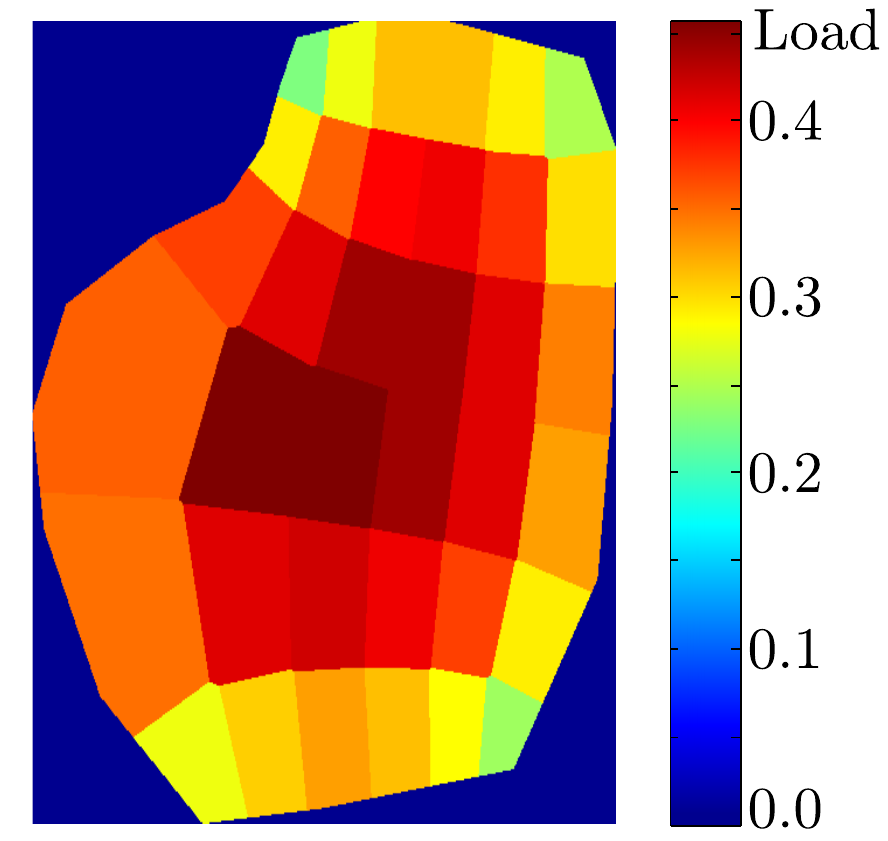}}	
					\subfloat[Spatial pattern: $\mathcal{A}_2$@hex]
	    		{\label{Fig:FIG12_7}
	    		\includegraphics[width = 0.241\textwidth]{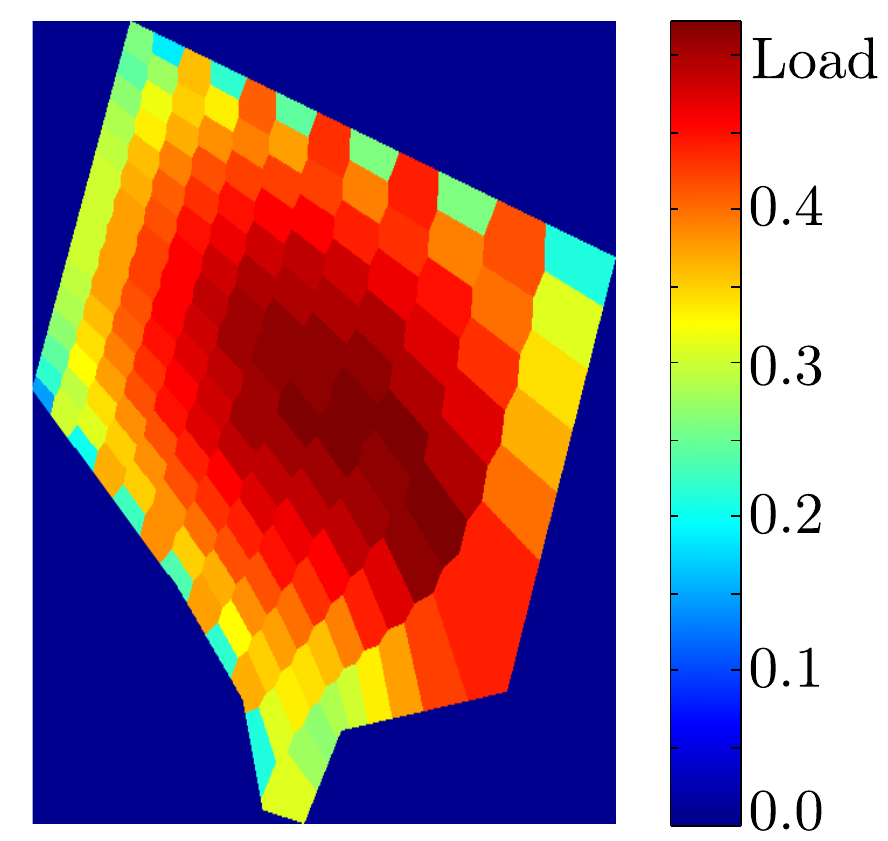}}	
					\subfloat[Spatial pattern: $\mathcal{A}_2$@rec]
	    		{\label{Fig:FIG12_8}
	    		\includegraphics[width = 0.241\textwidth]{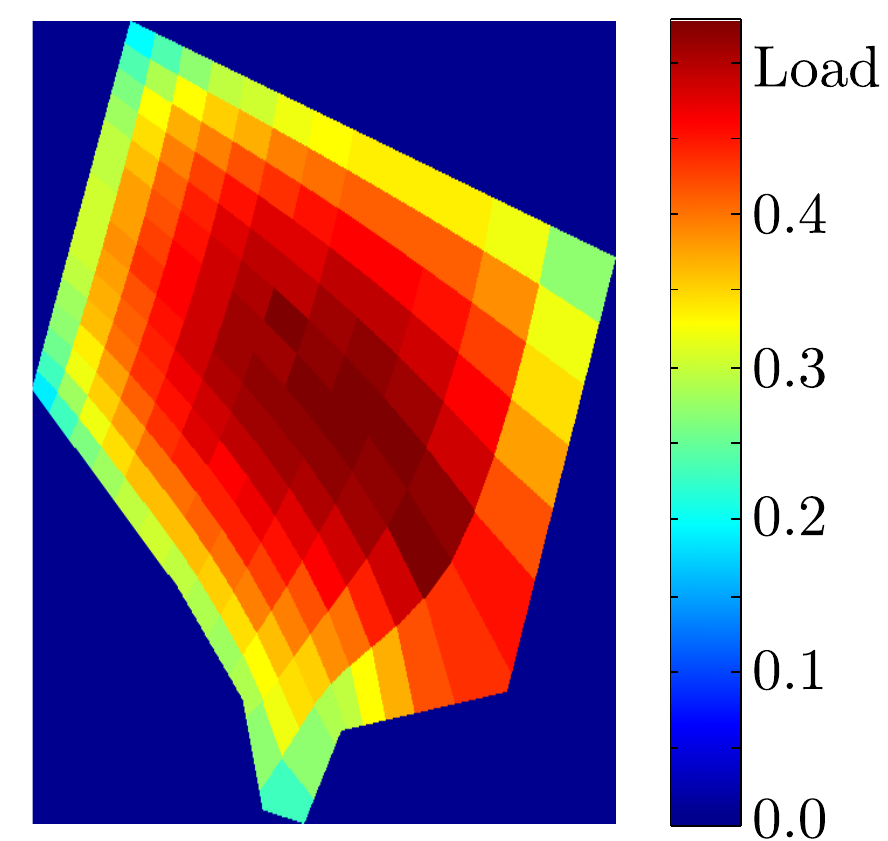}}	
	    		\vspace{-0.2cm}\caption{Numerical examples of load-coupling analysis: comparison of results between canonical and physical domains.}
	    		\label{Fig:FIG_12}    		
			\end{figure*}

			The analysis in the canonical domain allows for calculating the number of base stations ($L$) required to fulfill the service demand volume with a uniform load $\bar{\alpha}_{\text{c}}$. Indeed, such analysis depends exclusively on $V$ and it could be done in $\mathcal{R}$ without computing $F$. However, computing $F$ allows obtaining $F^{-1}$, and hence, here is when conformal mapping plays a key role. As indicated before, the idea is use the inverse mapping of $F$ (that makes $\delta$ regular), i.e., $F^{-1}$, to place the base stations in the physical domain given that $F$ captures the irregular nature of the spatial service demand distribution, see Figure~\ref{Fig:FRAMEWORK}. The site mappings $\mathcal{R}_1\leftrightarrow\mathcal{A}_1$ and $\mathcal{R}_2\leftrightarrow\mathcal{A}_2$ of our two examples for $L=36$ and $L=195$, respectively, are illustrated in Figure~\ref{Fig:FIG_SiteMapp}. Thus, Figures~\ref{Fig:FIG5_Phy02hex}~and~\ref{Fig:FIG5_Phy03hex} show the resulting topology if a hexagonal tiling is employed. As it can be seen, the aspect ratio $\frac{H}{W}$ of $\mathcal{R}$ cannot be fit with the conformal modules $m(Q)$ when regular hexagons are used, and hence, an intermediate linear transformation is required. The resulting rectangular tilings are shown in Figures~\ref{Fig:FIG5_Phy02sq}~and~\ref{Fig:FIG5_Phy03sq}. The figures also indicate the conformal modules, $m(Q)$, of the physical domains, and aspect ratios, $\frac{H}{W}$, of the canonical domains. \\
		\indent The corresponding load coupling analysis is shown in Figure~\ref{Fig:FIG_12}. Therein, Figures~\mbox{\ref{Fig:FIG12_1}-\ref{Fig:FIG12_4}} indicate the resulting load level of each cell in the canonical domain, with and without periodicity, and in the physical domain, without periodicity. It should be noted that, in all the cases, $\bar{\alpha}_{\text{c}}>\text{max}\{\bar{\alpha}_l\},\,\forall\,l\in[1,L]$ where $\bar{\alpha}_l$ refers to load in cell $l$ when the load coupling analysis is carried out in the canonical domain without periodicity (the dashed orange pattern). We note that, the load $\bar{\alpha}_{\text{c}}$ in the periodic canonical domain represents the worst case scenario in terms of intercell interference that flows over the domain boundary. In nonperiodic canonical and physical domains, the load in the cells close to domain boundary becomes smaller, which is also shown in the oscillating load pattern (dashed orange and green curves). 	
		The main interest is placed on the correlation between the load coupling analysis (without periodicity) in both domains: $\mathcal{R}$ and $\mathcal{A}$. That is, the correlation between the dashed-orange and solid-green patterns. \textsl{It has been observed that, the better the match between the aspect ratio of $\mathcal{R}$ and the conformal module of $\mathcal{A}$, the higher the correlation between both load patterns.} This has been verified in numerous numerical examples, which obviously cannot be presented herein due to space constraints. Thus, it can be concluded that, the existence of the aforementioned correlation between the load pattern in both domains makes evident the consistency of the proposed method. The load coupling correspondence between the domains is achieved because the spatial transformation preserves well the relative service demand share of each cell. In Figures~\ref{Fig:FIG12_2}~and~\ref{Fig:FIG12_4}, where the fit is perfect the correlation is almost perfect cell-by-cell. The periodic pattern is related with the fact that boundary cells receive, in general, less intercell interference as it is illustrated in Figures~\ref{Fig:FIG12_5}-\ref{Fig:FIG12_8}. Furthermore, in cases where the aspect ratio of $\mathcal{R}$ is clearly different from the conformal module of $\mathcal{A}$, see Figures~\ref{Fig:FIG5_Phy02hex},~\ref{Fig:FIG12_1} and~\ref{Fig:FIG5_Phy03hex},~\ref{Fig:FIG12_3}, the distances between average load levels in both domains are larger, although preserving the same periodic pattern cell-by-cell. Small differences are caused by the fact that the inverse conformal mapping ($F^{-1}$) does not guarantees perfect service demand sharing once the corresponding Voronoi regions are calculated, but good approximations of them. Another source of \textit{relative error}, in case of certain regular polygons, such as hexagons, comes from the fact that no periodicity can be assumed in the physical domain, and accordingly the service demand share of the boundary cells varies more significantly as it can be seen in Figure~\ref{Fig:FIG12_7}. This problem is obviously alleviated when rectangles are used as it is shown in Figure~\ref{Fig:FIG12_8}. From a practical point of view, the fact that $\bar{\alpha}_{\text{c}}$ is  (provided that conformal module is preserved in the canonical domain) above the actual load levels in the physical domain is quite advisable because it provides a natural protection against instantaneous variations with respect to average service demand values. In addition, if large scale networks are analyzed by parts, in a clustered fashion, it also provides a sort of wrap-around to account with the ICI coming from neighbor clusters. The resulting network topologies, which are deterministic in nature, can be used for planning and optimization of new or existing deployments, respectively. \\
		\indent Finally, it is worth saying that, the resulting dispersion in the load pattern (central cells tend to be more loaded) due to the lack of interference coming from neighbouring cluster can be reduced by increasing the noise power arbitrarily in cells on the area boundary. With such adjustment, the correlation will not be altered and the periodic patterns (orange and green) would tend to match with the red line, i.e., $\bar{\alpha}_{\text{c}}$.

		\section{Conclusions and Future Work}\label{Sec:ConclusionsFutWork}
This paper introduced a framework based on the idea of analyzing cellular networks through dual domains and spatial (conformal) transformations. The idea is novel and it provides a new approach to design, model, and analyze cellular network topologies \textbf{in a deterministic manner}, which is of great interest from both theoretical and practical point of view. Moreover, such network topologies are function of the service demand that has been characterized for the physical domain of interest, i.e., on the target coverage area. In this way, the service demand is modeled statistically, while the physical domain (the network coverage area) is represented deterministically by means of polygons.\\ 
\indent The numerical examples provide evidence that conformal mapping can be used as a tool to generate the required spatial transformations, while the load coupling analysis verifies a certain equivalence between the network topologies in the canonical and physical domain. Key conditions that must be fulfilled in order to improve the accuracy of the method have been also identified. Given that the work in the canonical domain is, in general, easier than the analysis of irregular topologies, the framework presented herein opens the road towards a novel analysis.
That is, the presented idea basically \textbf{opens a new research framework}, and consequently, the authors acknowledge that this is just the beginning and there is a plenty of future work ahead with several interesting open challenges. Among the most interesting ones are to postulate and verify some performance bounds, which are quite challenging in this context because some of the mathematical structures can only be dealt with through numerical methods. In addition, composite tessellations that can be used to model heterogeneous networks is, without a doubt, another interesting, yet challenging, research line that could be covered through the umbrella of idea presented herein. Finally, the rich framework of potential theory and boundary value problems will be of great help to understand or even device new spatial mappings that can also be used in the analysis of cellular networks through canonical domains. Given that conformal mapping and boundary value problems are strongly connected and have been used in other practical contexts, such as physics, this work also opens a promising line of research to formalize the notion of \textsl{network irregularity}, as the conformal mapping $F$ conveys the information about the structure and density of the network topology that is created. Current research efforts are aligned in that direction.

		\section*{Acknowledgment}\label{Sec:Ack}
		 This work was supported by Academy of Finland under grant $284811$.


\appendix
		\section*{Conformality and mapping composition}\label{App:ConfMapComp}
\subsection{The notion of conformality}\label{App:ConfMapComp_Conf}
				\begin{figure*}[t]
	    		\centering	  					
						\subfloat[conformal transformation]
	    		{\label{Fig:FIG_APP_CF1}
	    		\includegraphics[width = 0.32\textwidth]{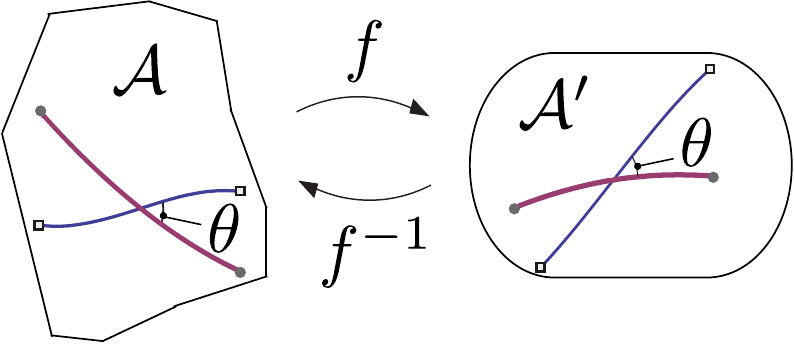}}	\hspace{0.16cm}		
															\subfloat[Complex curve and tangent]
	    		{\label{Fig:FIG_APP_CF2}
	    		\includegraphics[width = 0.25\textwidth]{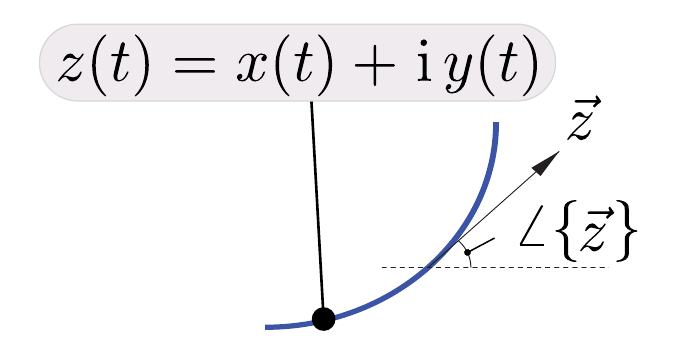}}\hspace{0.16cm}	
													\subfloat[The Riemann mapping theorem.]
	    		{\label{Fig:FIG_APP_RMT}
	    		\includegraphics[width = 0.28\textwidth]{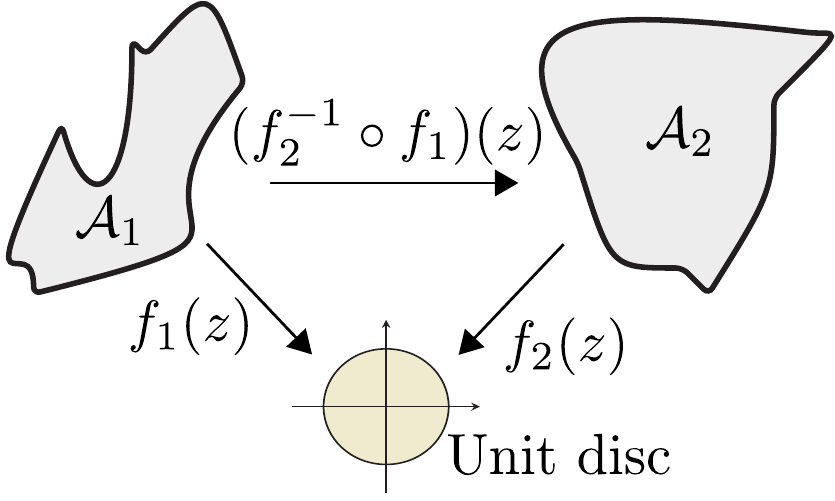}}	
	    		\vspace{-0.2cm}\caption{A \textit{conformal} transformation establishes a unique (one-to-one) mapping from points $a_0\in\mathcal{A}$ to other points $a_0^{\prime}\in\mathcal{A}^{\prime}$, where the function $f$ is \textit{analytic}. The points where $f$ is not analytic are called \textit{critical} points.}
	    		\label{Fig:FIG_APP}    		
			\end{figure*}	
A function $g:\spc{R}^n\rightarrow\spc{R}^n$ is called conformal if it preserves angles. What exactly does it mean? 
In the Euclidean norm, the angle between two vectors is defined by their dot product. However, most analytic maps are nonlinear, and so will not map vectors to vectors since they will typically map straight lines to curves. However, if we interpret `angle' to mean the angle between two curves, or more precisely, the angle between their tangent vectors at the point of intersection as illustrated in Figure~\ref{Fig:FIG_APP_CF1}, then it is possible to make sense of the conformality notion. Thus, in order to realize complex functions as conformal maps, it is needed to understand their effect on curves. In general, a curve in the complex plane is parametrized by a complex-valued function that depends on a real parameter $t$ as follows: $z(t) =x(t) \,+\, \text{i}\,y(t),~t\in\left[a,b\right]$. In particular, the tangent vector to the curve at the point $z$ can be identified with the complex number $\frac{\partial z}{\partial t} = \frac{\partial x}{\partial t} \,+\, \text{i}\,\frac{\partial y}{\partial t}$. Smoothness of the curve is guaranteed by the requirement that $\frac{\partial z}{\partial t}\neq 0$. 	When we interpret the curve as the motion of a particle in the complex plane, so that $z(t)$ is the position of the particle at time $t$, the tangent $\frac{\partial z}{\partial t}\neq 0$ represents its instantaneous velocity. The modulus of the tangent, $|\frac{\partial z}{\partial t}|=\sqrt{\left(\frac{\partial x}{\partial t}\right)^2\,+\,\left(\frac{\partial y}{\partial t}\right)^2}$, indicates the particle's speed, while its phase, $\angle \vec{z}$, measures the direction of motion, as prescribed by the angle that the curve makes with the horizontal, see Figure~\ref{Fig:FIG_APP_CF2}. The (signed) angle between two curves, $C_1$ and $C_2$, is defined as the angle between their tangents at the point of intersection $z=z_1(t_1)=z_2(t_2)$. If $\theta_1=\angle z_1(t_1)$ and $\theta_2=\angle z_2(t_2)$, then the angle $\theta$ at $z$ is $\theta=\theta_2-\theta_1$. Now, consider the effect of an analytic map $f(z)$. A curve $C$ parametrized by $z(t)$ will be mapped to a new curve $D = f(C)$, where $D(t)=f(z(t))$. The tangent to the image curve is related to that of the original curve by the chain rule: $\frac{\partial f(z)}{\partial t} = \frac{\partial f(z)}{\partial z}\frac{\partial z}{\partial t}$. 	Therefore, the effect of the analytic map on the tangent vector $\frac{\partial z}{\partial t}$ is to multiply it by the complex number $\frac{\partial f(z)}{\partial z}$. If the analytic map satisfies our key assumption $\frac{\partial f(z)}{\partial z}\neq0$, then $\frac{\partial z}{\partial t}\neq0$, and so the image curve is guaranteed to be smooth. Accordingly, $|\frac{\partial f(z)}{\partial t}| = |\frac{\partial f(z)}{\partial z}\frac{\partial z}{\partial t}|=|\frac{\partial f(z)}{\partial z}|\,|\frac{\partial z}{\partial t}|.$ Hence, the speed of motion along the new curve $f(z(t))$ is multiplied by a factor $|\frac{\partial f(z)}{\partial z}| > 0$. It is worth noting that the magnification factor $|\frac{\partial f(z)}{\partial z}|$ depends only upon the point $z$ and not how the curve passes through it. Indeed, all curves passing through the point $z$ are speeded up (or slowed down if $|\frac{\partial f(z)}{\partial z}| < 1$) by the same factor. Similarly, the angle that the new curve makes with the
horizontal is given by~$\angle \frac{\partial f(z)}{\partial t} = \angle \frac{\partial f(z)}{\partial z}\frac{\partial z}{\partial t}=\angle \frac{\partial f(z)}{\partial z}\,+\,\angle \frac{\partial z}{\partial t}.$ 	Therefore, the tangent angle of the curve is increased by an amount $\phi = \angle \frac{\partial f(z)}{\partial z}$, i.e., its tangent has been rotated through angle $\phi$. Again, the increase in tangent angle only depends on the point $z$ (and obviously, the transformation $f$), hence, all curves passing through $z$ are rotated by the same amount $\phi$. This immediately implies that the angle between any two curves is preserved.
\subsection{Existence and composition of conformal mappings}\label{App:ConfMapComp_MapComp}
		One of the hallmarks of conformal mapping is that one can assemble a large repertoire of \textit{complicated mappings} by simply composing elementary mappings. This relies on the fact that the composition of two complex analytic functions is also complex analytic. Hence, the composition of two conformal maps is also conformal. However, a fundamental question is: Is there, in fact, a conformal map $f(z)$ from an arbitrary domain $\mathcal{A}$ to the unit disk? The theoretical answer is the celebrated Riemann's mapping theorem. 		
	\textsl{	If $\mathcal{A}$ is any simply connected open subset, not equal to the entire complex plane, then there exists a one-to-one complex analytic map $f(z)$, satisfying the conformality condition $\frac{\partial f}{\partial z}\neq0$ for all $z\in\mathcal{A}$, that maps onto the unit disk $\spc{D} = {|z^{\prime}| < 1}$.} The proof can be found in~\cite[p.~221]{08:00078}. Thus, the Riemann's mapping theorem supports the notion of mapping composition as shown in Figure~\ref{Fig:FIG_APP_RMT}.


		\end{document}